# SAIL: Sound Abstract Interpreters with LLMs

QIUHAN GU, University of Illinois Urbana-Champaign, USA
AVALJOT SINGH, University of Illinois Urbana-Champaign, USA
GAGANDEEP SINGH, University of Illinois Urbana-Champaign, USA

How to construct globally sound abstract interpreters to safely approximate program behaviors remains a bottleneck in abstract interpretation. In this paper, we show the potential of using state-of-the-art LLMs to automate this tedious process. Focusing on the neural network verification area, we synthesize non-trivial sound abstract transformers across diverse abstract domains using LLMs to search within infinite space from scratch. We formalize the synthesis task as a constrained optimization problem, for which we design a novel mathematically grounded cost function that measures the degree of unsoundness of each generated candidate transformer, while enforcing hard syntactic and semantic validity constraints. Building on this formulation, we introduce Sail, a novel unified framework that combines model generation, syntactic and semantic validation, and cost-function-based refinement to synthesize globally sound abstract transformers. Evaluation results show that Sail not only matches the performance of manually designed transformers, but also is able to synthesize sound and high-precision transformers that do not exist in the literature for complex non-linear operators.

## 1 Introduction

Abstract Interpretation (AI) [13, 14, 67] is a popular framework for constructing automated program analyzers in different domains such as software [7, 15], machine learning [47, 62], and embedded systems [8, 33, 46]. These analyzers reason about an infinite number of program executions, producing invariants that can be used to prove properties such as safety, robustness, and stability. A key challenge in developing practical analyzers is obtaining *sound* abstract transformers, which overapproximate the effect of concrete program operations (e.g., assignments, conditionals). This is a tedious task and requires significant expert effort and heuristics. Indeed, considerable work exists on automatically synthesizing abstract interpreters [53], including symbolic synthesis techniques that derive constraints through formal reasoning [34, 35], deep-learning-based methods that infer transformer parameters from data and patterns [50], etc. Despite these advances, the synthesis process remains largely manual, domain-specific, computationally expensive, and often limited to obtaining sound transformers for individual operators. To the best of our knowledge, there exists no highly automated and efficient synthesis pipeline capable of generating abstract transformers that are both provably sound for all input abstract elements and generalizable across diverse operators and abstract domains.

Authors' Contact Information: Qiuhan Gu, University of Illinois Urbana-Champaign, Urbana, USA, qiuhang2@illinois.edu; Avaljot Singh, University of Illinois Urbana-Champaign, Urbana, USA, avaljot2@illinois.edu; Gagandeep Singh, University of Illinois Urbana-Champaign, Urbana, USA, ggnds@illinois.edu.

Recent advances in state-of-the-art large language models (LLMs) have transformed the way algorithms, code, and even scientific knowledge are generated [11, 17, 26, 29, 38, 39, 41, 55, 68, 70, 76]. Systems such as AlphaEvolve [49] and FunSearch [54] have demonstrated that LLMs can not only assist in coding but also evolve novel and high-performing algorithms through continuous evaluation and feedback. Motivated by these works, we investigate whether LLMs can synthesize provably sound abstract interpreters. Specifically, we focus on generating code describing the computations of sound abstract transformers operating over abstract elements. Leveraging LLMs to synthesize such abstract interpreters can improve automation and make the technology accessible to non-experts. However, this synthesis problem is more challenging for LLMs compared to those considered in the literature [4, 28, 31, 48], posing several non-trivial challenges.

*Challenge 1: How can we ensure the validity and global soundness of synthesized transformers?* Due to the hallucination of large language models [74], the generated code often contains syntactic or structural errors that make it invalid for soundness verification. Even if an LLM can occasionally produce syntactically correct code, ensuring global semantic soundness remains nontrivial. Our framework incorporates two checking modules: (1) a lightweight static validation frontend inspired by compiler design techniques to catch structural and typing errors, combined with a model repair agent to automatically suggest fixes; (2) a formal verification tool based on SMT solvers that certifies the transformer's soundness under all abstract elements. Together, these two components guarantee that only globally sound transformers are accepted.

*Challenge 2: How can an LLM effectively search within an infinite space?* Synthesizing a sound abstract transformer is fundamentally difficult because soundness must hold for all abstract elements, which form **an infinite search space, and each abstract element may abstract an infinite number of concrete points**. To overcome this, we formalize the synthesis process as a constrained optimization problem, for which we design a novel mathematically grounded cost function that measures the degree of unsoundness of each generated candidate transformer, while enforcing hard syntactic and semantic validity constraints as optimization constraints. This continuous formulation transforms synthesis from a binary pass/fail judgment into a guided optimization process, which iteratively refines valid but unsound candidates based on counterexamples and quantitative feedback.

*Challenge 3: How can we guarantee the convergence of the synthesis?* Synthesizing abstract interpreters within an infinite search space naturally raises the question of the theoretical feasibility of whether the search process can be guaranteed to converge to a sound transformer in finitely many steps. To address this, we formally define a novel refinement rule governing each synthesis step, and prove that under several easy-to-satisfy requirements, the iterative optimization process monotonically decreases the cost function and terminates within finite steps, ensuring theoretical convergence to a globally sound transformer.

To ground our study, we focus on the verification of deep neural networks (DNNs), a domain where abstract interpretation has emerged as a successful approach for proving model robustness and safety [59, 63]. DNN verification works with bounded polyhedral abstractions, making it a relatively easier problem to handle for LLMs as a case study for automated synthesis than analyzing programs, which often involves unbounded polyhedra.

**This Work**. We present Sail, **the first** general framework for synthesizing globally sound abstract interpreters with state-of-the-art LLMs. At its core, Sail formalizes transformer synthesis as an iterative optimization problem guided by a soundness-driven cost function. Sail can be combined with any verifier for the global soundness of abstract interpreters. In this paper, we instantiate Sail using ConstraintFlow [56], a framework which provides a simple and declarative DSL that encodes transformer logic as symbolic equations, a unified interface ProveSound [57] based on SMT solvers (Z3) for global soundness verification of abstract interpreters for DNNs, and a compiler backend [58] to transform the DSL-based transformers into executable programs. Beyond ConstraintFlow, Sail can also be instantiated using soundness verifiers implemented in more expressive proof assistants such as LEAN [5, 22] or Rocq [30, 32]. Given an operator and an abstract domain provided by the user as inputs, Sail repeatedly prompts an LLM to propose candidate transformers, checks the validation, fixes errors automatically with a separate model agent, verifies valid candidates against the soundness constraints, and computes a cost $\mathcal{L}$ that measures how far the candidate deviates from satisfying the soundness conditions. The feedback $\mathcal{L}$ serves as a continuous cost signal, along with the counterexamples, guiding subsequent synthesis rounds until a sound transformer is found. For efficiency, a trivially sound transformer will be returned if Sail does not converge within a fixed number of attempts. We manually provide specific fallbacks for certain monotonic functions, and use the $\top$ element of the abstract domain as the general fallback.

**Illustrative Example.** To provide an intuition for the design of the core cost function in Sail, consider synthesizing a globally sound abstract transformer for the ReLU activation in the Interval domain. For clarity, we omit DSL-specific syntax and focus on the semantic behavior of the transformer. ReLU activation is defined as $f(x) = max(0, x)$ concretely. Our goal is to synthesize a globally sound abstract transformer $F^\sharp$, where soundness requires that for *all* abstract inputs $[l, u]$, the output of $F^\sharp$ safely over-approximates $\max(0, x)$ for *all* $x \in [l, u]$. Example sound abstract transformers include $F^\sharp([l, u]) = [-\infty, \infty]$, $F^\sharp([l, u]) = [0, max(0, u)]$, or $F^\sharp([l, u]) = [0, 0]$ if $u \leq 0$, $[l, u]$ if $l \geq 0$, and $[0, u]$ otherwise, etc., all of which are globally sound but differ in precision. We employ few-shot prompting to bias the model away from trivial yet uninformative transformers (e.g., $F^\sharp([l, u]) = [-\infty, \infty]$), but the challenge remains to guide the model within an infinite search space. There have been various works that explored feedback-driven synthesis [34]. However, such feedback is typically symbolic and binary (i.e., fail/pass), making it insufficient to guide the synthesis procedure towards convergence. In contrast, we address this by introducing a fine-grained cost function that can quantitatively measure the degree of unsoundness of a candidate transformer by progressively analyzing abstract elements that violate the soundness condition down to their concrete states. For example, the candidate $F^\sharp([l, u]) = [0, 1]$ is locally sound when $l \leq u \leq 1$, but fails to approximate the output when the input interval lies partially or entirely outside $[-\infty, 1]$. In particular, consider one violating abstract input $[l, u] = [1, 3]$. Examining its concretization, we observe that concrete states such as $x = 2$ and $x = 3$ lead to outputs $f(x) > 1$, violating the constraint set by the upper bound. Numerically measuring the distance between the concrete outputs and the constraint provides a way to evaluate each candidate and quantify how far it deviates from satisfying the global soundness condition. We then take the worst-case deviation over all violating abstract elements as the cost. In practice, exhaustively enumerating all violating abstract elements and their corresponding concrete states is infeasible, as this space can be infinite. We instead adopt a sampling-based relaxation strategy coupled with a weight function to approximate the cost, ensuring instantiation feasibility. At each iteration, the model generates a batch of candidate transformers. If any candidate passes the soundness verifier, the process terminates with a sound transformer. Otherwise, the transformer candidate with the lowest cost, along with the corresponding counterexample, is then fed into the next iteration, guiding the model to correct previous errors.

**Main Contributions.** Our work makes the following contributions:

(1) An LLM-based synthesis framework. We design Sail, the first iterative synthesis system that couples LLM generation with symbolic verification to automate the synthesis of globally sound abstract transformers.
(2) Soundness-driven cost function. We introduce a novel soundness deviation metric, quantifying the degree of unsoundness and driving continuous refinement. With it, the pipeline either converges to a sound transformer or falls back to a trivially sound one after reaching the maximum number of attempts.

(3) Implementation and evaluation. We implement Sail on top of ConstraintFlow and demonstrate its ability to synthesize sound transformers across diverse operators and abstract domains. Our evaluation shows that Sail attains performance on par with handcrafted transformers for common non-linear operators, while further demonstrating the capability to efficiently synthesize novel and complex transformers with consistently high precision.

Beyond neural network certification, we further show that the principles underlying Sail can be extended to other research areas that require the automated construction of provably sound algorithms.

## 2 Background

### 2.1 Abstract Interpretation

Abstract interpretation [13] provides a mathematical foundation for sound reasoning about program behaviors by approximating all possible executions within a unified framework. It formalizes the principle of computing with abstractions rather than enumerating individual executions. A verifier based on abstract interpretation [21] reasons over two domains: the concrete domain $(C, \sqsubseteq_C)$, which represents the exact semantics of the system, and the abstract domain $(\mathcal{A}, \sqsubseteq_A)$, which encodes symbolic over-approximations of $C$. The two are connected by an abstraction function $\alpha : C \to \mathcal{A}$ and a concretization function $\gamma : \mathcal{A} \to C$. Soundness requires that for all $c \in C$, $c \sqsubseteq_C \gamma(\alpha(c))$, meaning that every concrete behavior is preserved within its abstraction. Within this framework, a program statement or neural-network operator can be modeled as a concrete transformer $F : C \to C$ and a corresponding abstract transformer $F^{\#} : \mathcal{A} \to \mathcal{A}$. The abstract transformer $F^{\#}$ is sound if and only if $F(\gamma(z)) \sqsubseteq_C \gamma(F^{\#}(z))$ for all $z \in \mathcal{A}$. Successive application of $F^{\#}$ over the structure of a program or network yields an over-approximation of all reachable states, which guarantees that any verified property holds for every concrete execution. Since abstract domains admit many incomparable abstractions, there is in general no single "best" sound transformer. Different transformers trade off precision and computational structure in domain-dependent ways. Therefore a key challenge in abstract interpretation is balancing precision and efficiency [19, 45]. A more complicated abstract domain, such as Polyhedra [64], often improves precision but is computationally expensive, while simpler domains such as Intervals [25] scale better but yield looser bounds. For neural networks, specialized domains such as DeepPoly [61], DeepZ [60], and CROWN [77] combine linear relaxations with neuron-wise constraints to achieve sound yet tractable over-approximations.

*Notation.* To avoid ambiguity and ensure notational clarity throughout the rest of this paper, we fix the following convention: bold symbols (e.g., $\mathbf{L}, \mathbf{U}$) denote vectors. We use $c \in C$ to denote a concrete element, $z \in \mathcal{A}$ to denote an abstract element. We use $\mathbf{x} \in \mathbb{R}^n$ to denote a concrete state, i.e., a complete assignment to all input variables of the program, whereas $x_i \in \mathbb{R}$ refers specifically to the value of the $i$-th variable within that state, $i \in [n]$. This distinction will be maintained consistently in all subsequent formal definitions and derivations.

### 2.2 DNN Certifier

A deep neural network (DNN) can be represented as a composition of affine and non-linear layers such as ReLU, Tanh, or MaxPool. Formally, a trained DNN defines a function $f : \mathbb{R}^m \to \mathbb{R}^n$. A verification property is described by a precondition $\varphi$, which denotes the admissible input region, and a postcondition $\psi$, which expresses the desired safety constraint on outputs. The goal of verification is to prove that $f(\varphi) \subseteq \psi$, meaning that no input $\mathbf{x} \in \varphi$ produces an unsafe output $f(\mathbf{x}) \notin \psi$. Abstract-interpretation-based verifiers [21, 63] compute a sound over-approximation of $f(\varphi)$. Starting from an abstract element $\alpha(\varphi)$, the verifier propagates it through the network using layer-wise abstract transformers and obtains an abstract output $g(\alpha(\varphi))$ such that $\gamma(g(\alpha(\varphi))) \supseteq f(\varphi)$. If $\gamma(g(\alpha(\varphi))) \subseteq \psi$, the property holds. Otherwise, the verifier may produce counterexamples or apply refinement strategies to reduce over-approximation. This layer-wise reasoning forms the basis of systems such as DeepPoly, CROWN, and ConstraintFlow. A DNN certifier integrates these verifiers into an end-to-end pipeline that formally proves safety, robustness, or other properties of networks. Certified training frameworks further embed verification into the learning process by shaping the loss function using verifier feedback, guiding the model toward provable robustness.

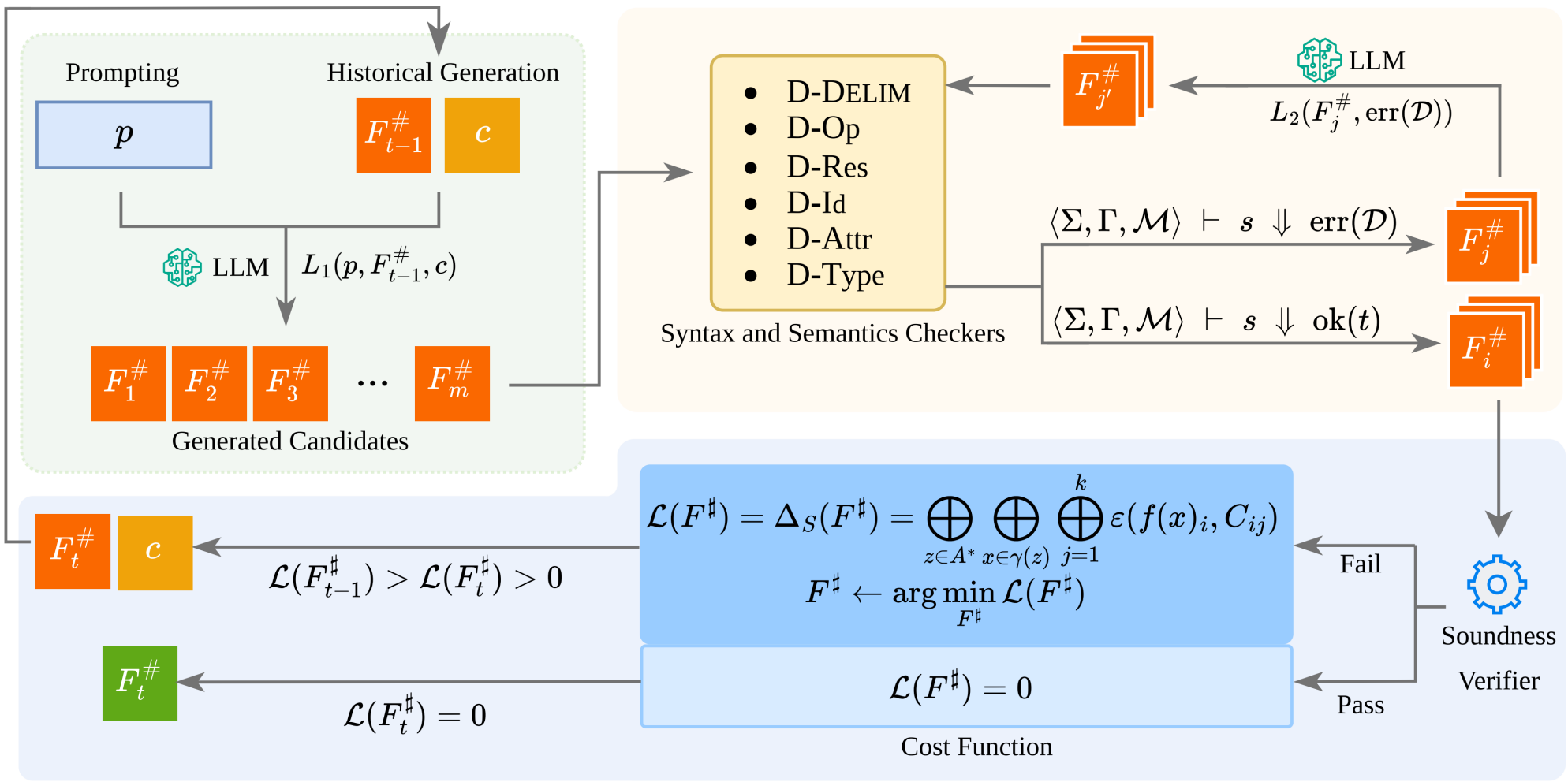

Fig. 1. The overview of SAIL. Given a prompt $p$, which specifies an operator, the domain specific language and abstract domain, and previous generation history, which includes the unsound transformer $F^{\sharp}_{t-1}$ and the counterexample $c$, the LLM $L_1$ proposes a set of candidates $\{F^{\sharp}_i, i \in [1, m]\}$, which will be validated by syntax and semantics checkers. Failing candidates trigger an automatic repair process based on another model agent $L_2$ until they pass the checkers. Valid candidates are verified by a soundness verifier and scored by a soundness-driven cost $\mathcal{L}(F^{\sharp})$. If their score is less than that of $F^{\sharp}_t$, then they will replace $F^{\sharp}_t$ as the current "best" unsound transformer in the prompt. This feedback loop transforms synthesis into an optimization process, refining candidates until $\mathcal{L}(F^{\sharp}) = 0$.

## 3 Overview

Fig. 1 presents the high-level idea of the constrained optimization process behind SAIL. Given an operator specification and an abstract domain specification in text as inputs, SAIL augments the prompt with DSL grammar and few-shot examples to construct the initial prompt $p$. SAIL searches for sound abstract transformers through an iterative process that combines generation, validation, and verification under the guidance of a soundness-oriented cost function $\mathcal{L}(F^{\sharp})$. While the current workflow assumes a predefined DSL, it remains fully extensible to other programming languages, as long as a corresponding validation module and a compatible soundness verifier are supplied.

A major difficulty in program synthesis with LLMs is that unconstrained LLM generation often produces code that is either syntactically invalid or semantically inconsistent. To mitigate this and improve the validity rate, SAIL generates multiple candidates in each pass. Each candidate is analyzed by a validator that parses it into an abstract syntax tree (AST) and verifies it against hard syntactic and semantic constraints. We provide general semantics for multiple syntax and semantic checks that can be applied in various programming languages in Appendix C. When errors appear, a dedicated repair model is invoked, which receives diagnostic messages and violating code region as feedback, and incrementally proposes corrections until the AST can be parsed and interpreted successfully, or until the maximum number of trials has been reached. The candidate will be discarded in the latter case. This automated repair loop is more efficient than the discard-and-retry strategy common in prior synthesis systems [3, 20, 65], enabling Sail to stabilize generation and preserve useful partial results, as shown in the ablation study in §5.4.

We verify all candidates that pass validation for soundness using a symbolic certifier. Instead of treating verification as a binary pass or fail decision, Sail evaluates each transformer using a novel cost function that quantifies its degree of soundness. The design of the cost function ensures that the cost of any sound transformer is 0. If that happens, then the transformer will be returned as the final result. If none of the candidates are sound, then the candidate with the lowest score is retained as the current "best" unsound transformer, and its associated counterexamples and score are incorporated into the next round of generation, functioning as a new and non-stochastic starting point in the search space and helping the model better understand the synthesis task. This feedback loop transforms synthesis into a continuous optimization process, iteratively refining candidates until a sound one is found. A trivially sound transformer will be returned as a fallback if no sound abstract transformer is found within the maximum number of attempts, ensuring a sound outcome of the overall procedure. We provide fallbacks based on endpoint-based bounds for certain monotonic functions, such as ReLU and HardSigmoid, and use the $\top$ element of the abstract domain as a general fallback. Details can be found in Appendix H. Empirically, we observe that Sail is able to successfully synthesize all transformers without using the fallback.

In summary, Sail enables a unified synthesis process that scales across diverse operators and abstract domains. The generated DSL-based transformers serve as symbolic specifications that are provably sound for all abstract inputs and can be effortlessly integrated into existing network certification frameworks through a compiler backend [58], surpassing purely mathematical formulations.

### 3.1 Illustrative Example

To illustrate the workflow of Sail, we demonstrate how it constructs a sound abstract transformer for the HardSigmoid activation for the popular DeepPoly [61] abstract domain based on an open-source LLM Llama4-Maverick, which provides free API access while exhibiting decent synthesis performance, sufficient to demonstrate the effectiveness of our framework. We choose HardSigmoid because it is non-trivial to handle, and there does not exist a globally sound DeepPoly transformer in the literature. This design choice prevents the language model from relying on memorized templates or prior retrievals, thereby exposing the genuine synthesizing capability of Sail. Formally, the HardSigmoid function is defined as a piecewise-linear function, as shown in the Fig. 2.

*3.1.1 Abstract Domain.* Next, we describe the DeepPoly domain, which associates two polyhedral and two interval constraints with each neuron, where a neuron represents the output value of a single computational node in the neural network. Formally, an abstract element is represented as $z = \langle \mathbf{l}, \mathbf{u}, \mathbf{L}, \mathbf{U} \rangle$. Here, $\mathbf{L}$ and $\mathbf{U}$ are vectors of affine functions over all neurons feeding into the current layer, with $L_i$ and $U_i$ representing the lower and upper polyhedral bounds of $x_i$ (where $x_i$ denotes the concrete value of the $i$-th neuron) and $l_i, u_i \in \mathbb{R}$ are the corresponding concrete lower and upper scalar bounds. The concretization is defined as $\gamma_n(z) = \{ \mathbf{x} \in \mathbb{R}^n \mid \forall i \in [n],\ l_i \leq x_i \leq u_i \wedge L_i \leq x_i \leq U_i \}$ DeepPoly is equipped with abstract transformers specifically tailored for verifying neural networks.

$$\text{HardSigmoid}(x) = \begin{cases} 0, & x \leq -3, \\ \dfrac{x+3}{6}, & -3 < x < 3, \\ 1, & x \geq 3. \end{cases}$$

(a) Piecewise definition.

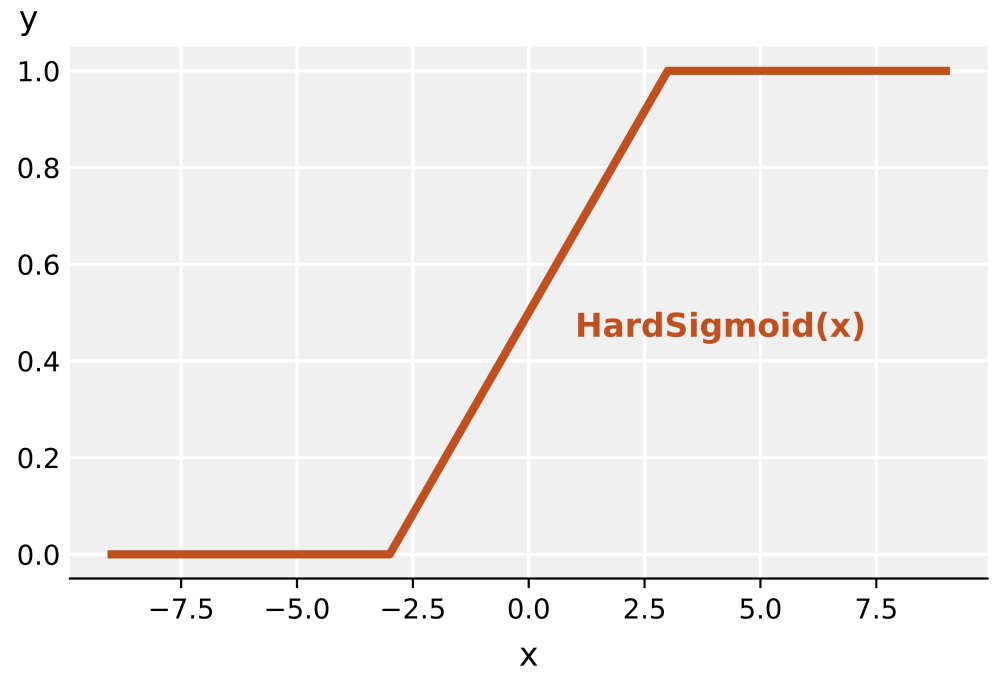

(b) Visualization of the function.

Fig. 2. Definition and visualization of the HardSigmoid activation. HardSigmoid linearly approximates the standard sigmoid between −3 and 3, and saturates at 0 and 1 outside this range.

*3.1.2 Domain-Specific Language (DSL).* We choose the domain-specific language and symbolic verification engine that are provided in ConstraintFlow [56] to support the synthesis process. Unlike general-purpose languages such as Python, the ConstraintFlow DSL emphasizes the mathematical specification of transformers while abstracting away implementation details, making it high-level and solver-friendly while remaining human-readable. Also the DSL is equipped with a soundness verification tool ProveSound built upon an SMT solver (Z3), which can automatically check the soundness of a given candidate transformer written in ConstraintFlow, either proving soundness for all abstract elements (instead of input-specific soundness) or generating a counterexample. ConstraintFlow DSL expresses abstract transformers as declarative equations relating input and output bounds. Each transformer specifies how an operator updates the lower and upper bounds $\langle \mathbf{l}, \mathbf{u}, \mathbf{L}, \mathbf{U} \rangle$ of the abstract element. Notably, while DeepPoly defines abstract transformers over the entire network state, including the preservation of bounds for all preceding neurons, ConstraintFlow simplifies this formulation by focusing only on the updated neurons. That is, instead of explicitly encoding how all previous neurons' bounds remain unchanged, the DSL abstracts away these preserved updates and specifies only the transformation relation between the input neuron $x_j$ and the output neuron $x_i$ affected by the current operator. This design yields a concise representation that is well-suited for LLM-based synthesis and symbolic verification.

*3.1.3 Iterative Synthesis.* We now demonstrate how Sail automatically synthesizes a sound transformer for the HardSigmoid operator within this setting.

*LLM Generation.* Each synthesis round begins with a structured prompt that encodes the operator specification, the semantics of the DeepPoly domain, and the grammar of the ConstraintFlow DSL, two-shot exemplars of verified transformers for related operators (Abs, Affine), and contextual feedback from previous iterations. Detailed prompt templates are provided in Appendix A. The large language model then produces a set of candidate transformers represented in DSL code, which can be categorized into one of three types: (i) syntactically or semantically invalid (e.g., unmatched parentheses or undefined metadata), (ii) unsound but syntactically and semantically valid, or (iii) valid and sound. Fig. 9 in Appendix B illustrates typical examples of each outcome.

*Validation and Repair.* After generation, each candidate undergoes a lightweight validation stage inspired by compiler frontends. Sail parses the candidate into an abstract syntax tree (AST) and applies hard-coded static checks for common structural and semantic errors, including:

(i) unmatched or missing delimiters such as parentheses and braces;

(ii) illegal keywords or illegal logical operators (e.g., using "&&" when ConstraintFlow only supports "`and`"; only provide one operand for the "and" operator);
(iii) improper use of reserved constants or keywords (e.g., naming a new function as "`transformer`", which is a reserved keyword).
(iv) undefined identifiers or invalid function invocations;
(v) malformed attribute calls and incorrect metadata indexing (e.g., using "." to access metadata instead of "[]"; using nested or numeric indices where symbolic ones (`l`, `u`, `z`) are expected);
(vi) type inconsistencies in arithmetic or element-wise operations (e.g., adding booleans to neurons);

These checks are detailed in Appendix C. When a violation is detected, the system invokes a dedicated repair agent. Detailed prompt templates are provided in Appendix A. If the error cannot be matched to any predefined rule, the agent receives a generic "`Unknown Error`" prompt and performs contextual repair until the AST passes validation or the maximum try limits are met.

*Soundness Verification and Cost Evaluation.* After a candidate passes static validation, it is submitted to the soundness verifier ProveSound. One of our key contributions is designing a novel cost function to quantify the degree of unsoundness, which captures how far an unsound candidate is from being sound. If the candidate is proved sound, then the cost function evaluates to 0 and the procedure terminates. Otherwise, the solver returns counterexamples that are used by the cost function to quantify the degree of unsoundness.

Designing such a cost function is highly non-trivial for two reasons. First, it is unclear how to quantify the deviation of an abstract element's concretization from its sound abstract enclosure in a mathematically meaningful way. Second, the soundness definition of an abstract transformer is expressed as a universal condition over all abstract elements [12, 23], which form an infinite set, making direct computation infeasible. To address these challenges, we begin by dissecting the definition of soundness itself. We progressively analyze each violating abstract element $z$, examining its concretization $\mathbf{x} \in \gamma(z)$, and relating each neuron $x_i$ within these concrete states to their abstract shape (or constraints) components $C_{ij}$. This stepwise reasoning reveals how deviations arise between the concrete and abstract semantics, enabling a mathematically grounded ideal formulation of the cost function that quantifies the extent of unsoundness. Since soundness is defined over an infinite set of abstract elements, we further introduce a relaxation strategy that approximates this ideal cost by sampling a finite subset and weighting them with an importance function $w(x_i)$ derived from the operator's gradient. Formally, the ideal cost function is given by

$$\mathcal{L}(F^\sharp) = \Delta_S(F^\sharp) = {}^{z}\bigoplus_{z \in A^*} {}^{\mathbf{x}}\bigoplus_{\mathbf{x} \in \gamma(z)} {}^{j}\bigoplus_{j=1}^{k} \varepsilon(f(\mathbf{x})_i, C_{ij}).$$

Detailed derivation can be found in §4. Here, $A^*$ denotes the set of all abstract elements for which the candidate transformer fails to satisfy $F(\gamma(z)) \sqsubseteq_C \gamma(F^\sharp(z))$. $z$ represents one abstract element from $A^*$, and $\mathbf{x} \in \gamma(z)$ corresponds to one of $z$'s concretizations, where $\mathbf{x}$ denotes the vector of neuron values representing the concrete network state. Let $x_i$ denote the value of the $i$-th neuron which will be updated in the state $\mathbf{x}$; $f(\mathbf{x})_i$ denotes the new value of $i$-th neuron after applying the concrete operator $f$ on $\mathbf{x}$ (e.g., the HardSigmoid activation). Since the operator updates only one neuron at a time while leaving others unchanged (i.e., $f(\mathbf{x})_j = x_j, j \neq i$), no soundness violation can occur in the unaffected neurons, making it safe to focus solely on the updated one. $C_{ij} \in \{l_i, u_i, L_i, U_i\}$ represents the $j$-th constraint derived from the $i$-th neuron's abstract shapes after applying the abstract transformer (e.g., interval or affine constraint). Each violation $\varepsilon(f(\mathbf{x})_i, C_{ij})$ quantifies how much the concrete output $f(\mathbf{x})_i$ falls outside its sound abstract enclosure, and the aggregation operator $\bigoplus$ accumulates these deviations into a single scalar measure of unsoundness. We differentiate aggregation operators based on their operands. ${}^{z}\bigoplus$, ${}^{\mathbf{x}}\bigoplus$, ${}^{j}\bigoplus$ represent aggregation operators operating on abstract elements, concrete states, and constraints, respectively. In order to ensure the convergence of the algorithm, each $\bigoplus$ is required to be a monotone, non-negative and bounded function satisfying the condition that: $\forall S_1, S_2$ as two of sets of operands, $S_1 \subseteq S_2 \implies \bigoplus S_1 \subseteq \bigoplus S_2$, $0 < \bigoplus S_1 < \infty, 0 < \bigoplus S_2 < \infty$. $\bigoplus S = 0$ if and only if $S = \emptyset$, meaning there is no soundness violations. Typical instantiations include maximum and mean. In practice, we use maximization as the universal aggregation operator. In the implementation, we apply an approximation strategy combining sampling and weight function. Details can be found in §4.5. Let $\langle \mathbf{l}, \mathbf{u}, \mathbf{L}, \mathbf{U} \rangle$ denote the input abstract element, and $\langle \mathbf{l}', \mathbf{u}', \mathbf{L}', \mathbf{U}' \rangle$ the corresponding output element. Then, based on DeepPoly, the cost function is approximated as:

$$\begin{aligned}\mathcal{L}(F^\sharp) = \widetilde{\Delta_S}(F^\sharp) &= \max_{z \in A^*} \max_{\mathbf{x} \in \gamma_{\text{sample}}(z)} w(x_i) \cdot \max_{j=1}^{4} \varepsilon(f(\mathbf{x})_i, C_{ij}) \\ &= \max_{z \in A^*} \max_{\mathbf{x} \in \gamma_{\text{sample}}(z)} w(x_i) \cdot \Big( \max(0, f(\mathbf{x})_i - u'_i) + \max(0, l'_i - f(\mathbf{x})_i) \\ &\qquad + \max(0, f(\mathbf{x})_i - U'_i) + \max(0, L'_i - f(\mathbf{x})_i) \Big)\end{aligned}$$

where $\gamma_{\text{sample}}(z)$ represents a sampled subset of $\gamma(z)$. $w(x_i)$ denotes the weight function, $w(x_i) = \frac{\phi(f, x_i)}{\sum_{x' \in \gamma_{\text{sample}}(z)} \phi(f, x'_i)}$, $\phi(f, x_i) = \log\left(1 + \exp(\|\nabla_{x_i} f(x_i)\|)\right)$. This design of the weight function leverages the gradient magnitude of $f$ to prioritize semantically critical configurations, for instance, around $x = 3$ in the HardSigmoid function, where the gradient is high and transformers are more prone to errors. The softplus transformation in the weight function avoids vanishing contributions in flat regions. By pairing sampling with a weight function, we ensure that the finite-sample approximation remains representative of the full sampling space.

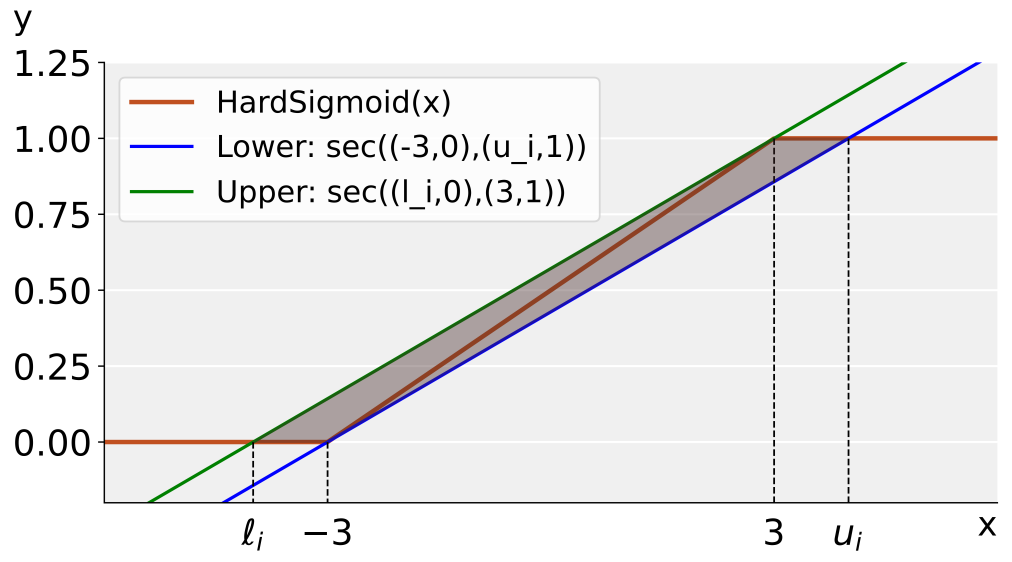

Fig. 3. Correct polyhedra bounds for HardSigmoid transformer on interval $[l_i, u_i]$ when $l_i < -3 < 3 < u_i$. The shaded areas highlight the safely approximated region.

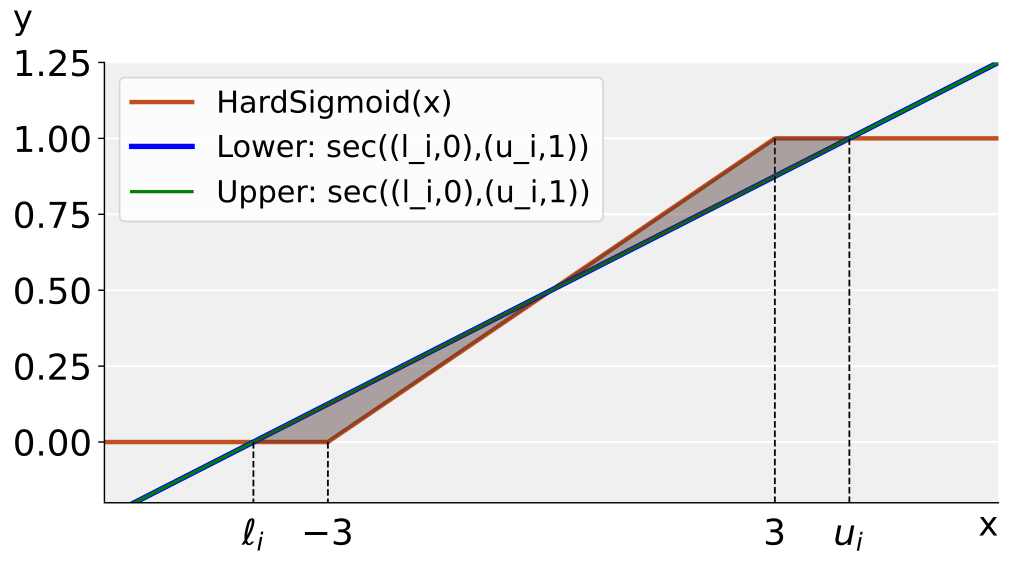

Fig. 4. Incorrect polyhedra bounds for HardSigmoid transformer on interval $[l_i, u_i]$, $l_i < -3 < 3 < u_i$. The shaded areas highlight the regions that are not safely approximated and violate the soundness condition.

We consider the most challenging case in synthesizing the HardSigmoid transformer when the input interval spans the interval $(-3, 3)$, where the function switches between the linear and saturated region, i.e., $l_i < -3 < 3 < u_i$. In this mixed case, the operator alternates between the saturated zones ($(-\infty, -3)$ and $(3, +\infty)$)) and the central linear region, yielding a non-convex shape that must be over-approximated by two affine bounds and the interval bounds. Since the HardSigmoid function is monotonic, the interval bounds are easy to get, while two distinct affine relaxations are hard to generate. As shown in the Fig. 4, an unsound candidate generated by Llama4-Maverick produces overlapping affine bounds, which corresponds to the secant line connecting $(l_i, 0)$ and $(u_i, 1)$. In this case, the unsound transformer $F_0^\sharp$ is then defined as:

$$F_0^\sharp(\langle l_i, u_i, L_i, U_i \rangle) = \langle l'_i, u'_i, L'_i, U'_i \rangle = \langle 0, 1, \frac{1-0}{u_i - l_i}(x_i - l_i), \frac{1-0}{u_i - l_i}(x_i - l_i) \rangle,$$

when $l_i < -3 < 3 < u_i$, which under-approximates part of the nonlinear curve and violates the soundness condition. The shaded regions in the Fig. 4 highlight these violating areas, compared to the sound approximation shown in Fig. 3.

To compute the cost function in this case, we first obtain counterexamples that violate the soundness constraints from the SMT solver, such as $z_1 = \langle(\ldots,-4,\ldots),(\ldots,4,\ldots),\mathbf{L},\mathbf{U}\rangle, z_2 = \langle(\ldots,-5,\ldots),(\ldots,4,\ldots),\mathbf{L},\mathbf{U}\rangle, z_3 = \langle(\ldots,-5,\ldots),(\ldots,5,\ldots),\mathbf{L},\mathbf{U}\rangle$, etc. Here, we only focus on the lower and upper bounds of the $i$-th neuron of each abstract element, i.e., $l_i$ and $u_i$, since the affine bounds of the input elements are irrelevant when evaluating the cost function, and the unchanged neurons would not affect soundness, as mentioned before. Take $z_1$ as an instance, we can have $x_i \in \{-4,-3,-2,-1,0,1,2,3,4\}$, where $\mathbf{x} \in \gamma_{sample}(z_1)$. Since the interval bounds $l_i' = 0$ and $u_i' = 1$ do not contribute to violations, the cost arises solely from the deviation between $f(\mathbf{x})_i$, i.e., $HardSigmoid(x_i)$, and the linear relaxation. The incorrect affine relaxation in this case used by the unsound transformer, i.e., $U_i' = L_i' = \frac{1}{8}x_i + \frac{1}{2}$, yields the violation term at each sampled point:

$$\begin{aligned}\varepsilon(f(\mathbf{x})_i, C_{ij}) &= \max(0, HardSigmoid(x_i) - U_i') + \max(0, L' - HardSigmoid(x_i)) \\ &= \left| HardSigmoid(x_i) - (\tfrac{1}{8}x_i + \tfrac{1}{2}) \right|.\end{aligned}$$

Evaluating over sampled $\{x_i\}$ yields $\varepsilon = \{0, 0.125, 0.0833, 0.0417, 0, 0.0417, 0.0833, 0.125, 0\}$ respectively. Each $\varepsilon(f(\mathbf{x})_i, C_{ij})$ is then multiplied by its corresponding weight, and the maximum weighted violation is taken as the final cost, i.e., $\mathcal{L}_{z_i}(F_0^\sharp) = \max_{\mathbf{x}\in\gamma_{\text{sample}}(z_1)} w(x_i)\,\varepsilon(x_i) \approx 0.11390.125 \approx 0.01424$, when $x_i = \pm 3$. Similarly, other abstract elements such as $z_2$ and $z_3$ are processed in the same way. We then accumulate their contributions to obtain $\mathcal{L}(F_0^\sharp)$:

$$\mathcal{L}(F_0^\sharp) = \max(\mathcal{L}_{z_i}(F_0^\sharp), \mathcal{L}_{z_2}(F_0^\sharp), \mathcal{L}_{z_3}(F_0^\sharp)) = \max(0.01424, 0.0230, 0.01895) = 0.0230.$$

In each iteration, when no sound transformer is obtained, we select from all candidates the one with the smallest cost value as the current "best" unsound transformer. This transformer, together with its counterexamples and cost score, is merged into the next prompt to guide subsequent iterations, encouraging the model to learn from prior mistakes and produce improved candidates. If the next round still fails to yield a sound transformer, we again identify the candidate whose cost is lower than the current "best" unsound one and feed it into the next iteration as the updated unsound transformer. A sound fallback transformer $F^\sharp(\langle l_i, u_i, L_i, U_i\rangle) = \langle 0, 1, 0, 1\rangle$ is provided if the synthesis fails within a fixed number of attempts. While in practice, Sail is able to converge to a more precise transformer that uses correct secant lines as the upper and lower polyhedral bounds, resulting in significantly tighter abstractions. The complete synthesizing process in this case is shown in the Fig. 6c.

## 3.2 Implicit Effects on Precision

In addition to soundness, precision is another metric typically considered when evaluating an abstract transformer [13, 34], reflecting the tightness of the abstraction. A more precise abstract transformer yields a smaller over-approximation gap between the abstract output $\gamma(F^\sharp(z))$ and the concrete output set $F(\gamma(z))$, thus enabling more properties to be successfully verified. While our framework formally guarantees soundness of the synthesized transformers, we empirically observe that these transformers often exhibit decent precision as well. We provide an intuitive explanation to justify this behavior.

The prompt to LLMs incorporates both few-shot examples that include one illustrating a precise transformer with case distinctions, and explicit DSL constructions for some common relaxation patterns, such as tangent and secant lines, as shown in Appendix A. These design choices introduce an inductive bias in synthesis that guides the model toward well-structured bounds and discourages trivial yet overly loose relaxations(e.g., $[-\infty, \infty]$). Based on this, precision is furthermore implicitly enforced by the cost function, as precision and soundness are not entirely independent in the early stage of the search. Candidates that appear tight are often not yet sound over-approximations, and thus verified as unsound. For example, incorrect case distinctions tend to incur violations near transition points of the concrete function, which are then penalized more heavily by the cost function, as it leverages a weight function that is based on the gradient magnitude of the concrete function to prioritize violations near regions where the function changes more sharply (e.g., 0 for the absolute value function). Also, affine bounds with misaligned slopes or intercepts may fail to enclose certain concrete outputs, as shown in the Fig. 4, leading to soundness violations that incur a non-zero cost. As these violations are progressively corrected through cost-guided refinement, the candidate may move from a tight but unsound approximation toward one that is both sound and precise. This also explains why the synthesized transformers produce tighter bounds than the trivially sound fallback, which either returns the top element or relies solely on scalar bounds.

## 4 Formalizing LLM-Guided Synthesis

### 4.1 Abstract Interpretation and Abstract Transformers

We consider the setting of numerical abstract interpretation. The concrete and abstract domains are denoted as $C$ and $\mathcal{A}$ respectively. Let $\mathbb{P}_C = (\mathcal{P}(\mathbb{R}^n), \sqsubseteq_C)$ be the poset on the power set of concrete states where $\sqsubseteq_C$ is subset inclusion. Let $\mathbb{P}_A = (\mathcal{A}, \sqsubseteq_A)$ be the poset on the abstract elements. The concretization function $\gamma : \mathcal{A} \to \mathcal{P}(\mathbb{R}^n)$ maps an abstract element to a set of concrete elements.

*Sound Abstract Transformer.* Let $f : \mathbb{R}^n \to \mathbb{R}^n$ be a concrete function. Its corresponding concrete transformer $F : \mathcal{P}(\mathbb{R}^n) \to \mathcal{P}(\mathbb{R}^n)$ is defined as: $c \in \mathcal{P}(\mathbb{R}^n),\ F(c) := \{f(\mathbf{x}) \mid \mathbf{x} \in c\}$. Given an abstract transformer $F^\sharp : \mathcal{A} \to \mathcal{A}$, we say $F^\sharp$ is *globally sound* if it overapproximates the concrete semantics for all abstract elements, i.e., $\forall z \in \mathcal{A},\ F(\gamma(z)) \sqsubseteq_C \gamma(F^\sharp(z))$. That is, applying the approximation $F^\sharp$ on any abstract element $z$, and then obtaining the set of concrete values corresponding to the result must include more states than first concretizing the abstract element and then applying the concrete transformer $F$.

### 4.2 Unsoundness Deviation and Metrics

Based on the definition of soundness, an abstract transformer $F^\sharp$ is *unsound* iff $\exists z \in \mathcal{A},\ F(\gamma(z)) \not\sqsubseteq_C \gamma(F^\sharp(z))$. By the definition of subset inclusion, this is equivalent to: $\exists z \in \mathcal{A},\ \exists \mathbf{y} \in \mathbb{R}^n,\ \mathbf{y} \in F(\gamma(z)) \wedge \mathbf{y} \notin \gamma(F^\sharp(z))$. From the definition of the concrete transformer $F$, we have $F(\gamma(z)) = \{f(\mathbf{x}) \mid \mathbf{x} \in \gamma(z)\}$. Thus, for any $\mathbf{y} \in F(\gamma(z))$, there exists some $\mathbf{x} \in \gamma(z)$ such that $\mathbf{y} = f(\mathbf{x})$. Substituting this into the previous expression gives: $\exists z \in A,\ \exists \mathbf{x} \in \mathbb{R}^n,\ f(\mathbf{x}) \in F(\gamma(z)) \wedge f(\mathbf{x}) \notin \gamma(F^\sharp(z))$. Since $f(\mathbf{x}) \in F(\gamma(z))$ whenever $x \in \gamma(z)$, this simplifies to: $\exists z \in A,\ \exists \mathbf{x} \in \mathbb{R}^n,\ x \in \gamma(z) \wedge f(\mathbf{x}) \notin \gamma(F^\sharp(z))$. That is, the abstract transformer $F^\sharp$ is unsound if there exists some abstract element $z$ whose concretization contains at least one state $\mathbf{x}$ such that $f(\mathbf{x})$ is not soundly captured by the concretization of the abstract output $F^\sharp(z)$. We collect all such violating abstract elements to get the *counterexample set*:

$$A^* := \left\{ z \in \mathcal{A} \mid \exists \mathbf{x} \in \gamma(z),\ f(\mathbf{x}) \notin \gamma(F^\sharp(z)) \right\}.$$

Notably, $A^*$ can be an infinite set. $A^*$ will be an empty set when the abstract transformer is sound.

To quantify how severely an abstract transformer $F^\sharp$ deviates from a sound one, we define a deviation metric $\Delta_S$ that aggregates violations across all elements in $A^*$ to quantify the extent to which the abstract transformer $F^\sharp$ fails to be sound:

$$\Delta_S(F^\sharp) := {}^{z}\bigoplus_{z \in A^*} \nu(z) \tag{1}$$

where $\nu(z) \in \mathbb{R}_{\geq 0}$ captures how much $F^\sharp$ fails to satisfy the soundness property with respect to the abstract element $z$. Here, ${}^{z}\bigoplus$ denotes a generic aggregation operator that combines the local violation measures across abstract elements. To ensure that $\Delta_S(F^\sharp)$ defines a well-behaved deviation measure, the aggregation operator ${}^{z}\bigoplus$ is required to be (1) monotone, (2) non-negative and (3) bounded. Formally $\forall S_1, S_2, S_1 \subseteq S_2 \implies {}^{z}\bigoplus S_1 \subseteq {}^{z}\bigoplus S_2$, $0 < {}^{z}\bigoplus S_1 < \infty, 0 < {}^{z}\bigoplus S_2 < \infty$. ${}^{z}\bigoplus S = 0$ if and only if $S = \emptyset$. Typical instantiations include maximum and mean.

A natural question arises: how should one define $\nu(z)$? A naive approach is to assign $\nu(z) = 1$, i.e., uniformly weighting each abstract element, so that $\Delta_S$ reduces to counting the total number of unsound abstract elements. However, this strategy is unsatisfactory for two reasons. First, the set $A^*$ may be infinite, which renders such counting intractable. Second, a uniform assignment fails to capture the heterogeneous contributions of different abstract elements to unsoundness. For example, some $z \in A^*$ may correspond to a violation caused by only a single concrete point, whereas others may exhibit violations across almost their entire concretization. Assigning them the same weight therefore discards crucial information about the relative severity of their contributions to unsoundness. To better capture the severity of soundness failure, we define $\nu(z)$ quantitatively by aggregating individual pointwise violations over all $\mathbf{x} \in \gamma(z)$. This enables a more fine-grained view of transformer quality and supports optimization-based repair strategies. At the same time, such a quantitative formulation enables approximating $\Delta_S$ through sampling in the future when needed. We decompose each abstract-level violation $\nu(z)$ into its underlying pointwise contributions by accumulating the violations from each $\mathbf{x} \in \gamma(z)$: $\nu(z) := {}^{\mathbf{x}}\bigoplus_{\mathbf{x}\in\gamma(z)} \delta(f(\mathbf{x}), \gamma(F^\sharp(z)))$. Here, ${}^{\mathbf{x}}\bigoplus$ is the aggregation operator working on concrete states $\mathbf{x}$, sharing the same three properties as ${}^{z}\bigoplus$. The pointwise metric $\delta(f(\mathbf{x}), \gamma(F^\sharp(z))) \in \mathbb{R}_{\geq 0}$ quantifies the degree to which the concrete output $f(\mathbf{x})$ lies outside the abstract output region $\gamma(F^\sharp(z))$. Formally, we require $\delta$ to satisfy the following property $\mathscr{P}$:

$$\delta(f(\mathbf{x}), \gamma(F^\sharp(z))) = \begin{cases} 0 & \text{if } f(\mathbf{x}) \in \gamma(F^\sharp(z)) \\ > 0 & \text{if } f(\mathbf{x}) \notin \gamma(F^\sharp(z)) \end{cases}$$

That is, $\delta$ evaluates to zero if all concrete outputs are captured by the abstract output, indicating no contribution to unsoundness. It returns a positive value if the output violates the soundness condition. Therefore, the total deviation becomes:

$$\Delta_S(F^\sharp) = {}^{z}\bigoplus_{z\in A^*} {}^{\mathbf{x}}\bigoplus_{\mathbf{x}\in\gamma(z)} \delta(f(\mathbf{x}), \gamma(F^\sharp(z))), \tag{2}$$

which provides a structured and quantifiable measure of how far $F^\sharp$ deviates from being a sound abstract transformer.

*Shape-aware deviation.* We define $n$ as the number of variables in the concrete domain captured by $z$. In the general case, these variables are not independent: constraints may relate multiple variables simultaneously, reflecting dependencies across the program state. Nevertheless, by applying Fourier–Motzkin Elimination(FME) [16], we can always rewrite such relational constraints into multiple constraints expressed with respect to a particular variable of interest. Thus, $F^\sharp(z)_i$ will have $n$ dimensions, each dimension encoding the set of constraints associated with the corresponding concrete variable: $F^\sharp(z)_i = (C_{i1}, C_{i2}, \ldots, C_{ik}), \quad i \in \{1, \ldots, n\}$, where each $C_{ij}$ denotes a scalar or affine constraint on the $i$-th variable obtained via FME, and $k$ denotes the number of constraints that can be derived. Given this componentwise interpretation, the total deviation $\Delta_S$ can be further decomposed into a nested aggregation over the abstract element $z$, its concretization $\mathbf{x} \in \gamma(z)$, the updated variable index $i$, and the $k$ constraints associated with that variable:

$$\Delta_S(F^\sharp) = {}^{z}\bigoplus_{z\in A^*} {}^{\mathbf{x}}\bigoplus_{\mathbf{x}\in\gamma(z)} {}^{i}\bigoplus_{i=1}^{n} {}^{j}\bigoplus_{j=1}^{k} \varepsilon\big(f(\mathbf{x})_i, C_{ij}\big). \tag{3}$$

where ${}^{i}\bigoplus, {}^{j}\bigoplus$ are the aggregation operators operating on neurons and constraints, sharing the same properties as ${}^{z}\bigoplus, {}^{\mathbf{x}}\bigoplus$. Consistent with the global deviation measure $\delta$, the shape-level violation function $\varepsilon$ is required to satisfy an analogous property $\mathscr{P}$:

$$\varepsilon\big(f(\mathbf{x})_i, C_{ij}\big) = \begin{cases} 0 & \text{if } f(\mathbf{x})_i \models C_{ij}, \\ > 0 & \text{if } f(\mathbf{x})_i \not\models C_{ij}, \end{cases}$$

where $f(\mathbf{x})_i \models C_{ij}$ denotes that the $i$-th component of $f(\mathbf{x})$ satisfies the $j$-th constraint $C_{ij}$.

*Single assignment function.* In many practical scenarios, the function $f$ is single-assignment, that is, it updates only one of the $n$ program variables while leaving the others unchanged. Formally, let $i$ denote the index of the updated variable, $f(\mathbf{x})_{i'} = x_{i'}, \forall i' \neq i$. Then we will have $f(v)_{i'} = x_{i'} \in \gamma(z)_{i'} = \gamma(F^\sharp(z))_{i'}$, i.e., the value of $f(\mathbf{x})_{i'}$ remains within the abstract output region. Hence, the shape-level violation for these variables vanishes, i.e., $\varepsilon(f(\mathbf{x})_{i'}, C_{i'j}) = 0$ for all $j = 1, \ldots, k$ and all $i' \neq i$. Therefore, the decomposition in Equation 3 simplifies to:

$$\Delta_S(F^\sharp) = {}^{z}\bigoplus_{z \in A^*} {}^{\mathbf{x}}\bigoplus_{\mathbf{x} \in \gamma(z)} {}^{j}\bigoplus_{j=1}^{k} \varepsilon\big(f(\mathbf{x})_i, C_{ij}\big). \tag{4}$$

where $i$ is the unique updated variable. The implementation of $\varepsilon$ naturally depends on the representation of the constraint $C_{ij}$, which in turn is determined by the chosen abstract domain. To systematically capture the structure of $C_{ij}$, we define them as below. We denote the neuron value by $y \in \mathbb{R}$ for clarity.

$$\begin{aligned} \langle \mathit{Constraint} \rangle \quad &::= \quad \langle \mathit{ScalarBound} \rangle \quad | \quad \langle \mathit{AffineBound} \rangle \\ \langle \mathit{ScalarBound} \rangle \quad &::= \quad y \leq c \quad | \quad y \geq c \\ \langle \mathit{AffineBound} \rangle \quad &::= \quad y \leq a_0 + \textstyle\sum_{i=1}^{n} a_i x_i \quad | \quad y \geq a_0 + \textstyle\sum_{i=1}^{n} a_i x_i \end{aligned}$$

Here, $c \in \mathbb{R}$ denotes a concrete scalar bound, and $a_0, a_1, \ldots, a_n \in \mathbb{R}$ are the coefficients of an affine function over symbolic variables $x_1, x_2, \ldots, x_n$. This definition captures both concrete constant bounds and symbolic affine expressions over input variables. By combining these two types of bounds, we can flexibly express a wide range of abstract domains. For instance, in the Interval domain, each constraint is defined by the scalar bound; In the Polyhedra domain, a constraint is represented by the affine bound, specifying a linear constraint on inputs; In the DeepPoly domain, each abstract shape is encoded with a pair of scalar bounds and a pair of affine bounds (one upper and one lower), allowing tighter symbolic enclosures for neural network verification, leading to the corresponding constraints taking the form of either scalar bounds or affine bounds:

(1) **Scalar Bound**. We use $m \in \mathbb{R}$ to denote the neuron value be evaluated (e.g., $f(\mathbf{x})_i$ as we discussed before). When the constraint is a scalar bound of the form $y \leq c$ or $y \geq c$, the violation is defined as the one-sided distance from the evaluated point $m$ to the feasible region: $\varepsilon(m,\ y \leq c) := \max(0,\ m - c), \varepsilon(m,\ y \geq c) := \max(0,\ c - m)$.
(2) **Affine Bound**. When the constraint is an affine bound $y \leq a_0 + \sum_{i=1}^{n} a_i x_i$ or $y \geq a_0 + \sum_{i=1}^{n} a_i x_i$, the violation is: $\varepsilon(m,\ y \leq a_0 + \sum a_i x_i) := \max(0,\ m - (a_0 + \sum a_i x_i)), \varepsilon(m,\ y \geq a_0 + \sum a_i x_i) := \max(0,\ (a_0 + \sum a_i x_i) - m)$.

## 4.3 Constrained Optimization Problem

To address the problem of synthesizing sound abstract transformers, we formalize the synthesis process as an LLMs-based constrained optimization problem that minimizes a quantitative cost function $\mathcal{L}(F^\sharp)$ subject to hard syntactic and semantic validity constraints.

*Search Space.* Let $\mathcal{H}$ denote the set of all candidate abstract transformers $F^\sharp$. Within $\mathcal{H}$, we define three disjoint subsets that partition the space according to validity and soundness:

$$\mathcal{V} := \{\, F^\sharp \in \mathcal{H} \mid F^\sharp \text{ is syntactically and semantically valid} \,\},$$
$$\mathcal{G} := \left\{ F^\sharp \in \mathcal{V} \;\middle|\; \forall z \in A,\ F(\gamma(z)) \subseteq \gamma(F^\sharp(z)) \right\}, \mathcal{U} := \mathcal{V} \setminus \mathcal{G}.$$

Here, $\mathcal{V}$ contains all valid transformers, $\mathcal{G} \subseteq \mathcal{V}$ contains all valid and sound transformers, and $\mathcal{U}$ contains all valid but unsound transformers that violate the soundness condition for some $z \in A$.

*LLMs Generation.* Given a prompt $p$, the large language model acts as a stochastic generation operator $\Pi_{\text{LLM}}$ that produces a batch of candidate abstract transformers:

$$\{F^\sharp_{cand}\} = \Pi_{\text{LLM}}(p), \quad \text{where} \quad \Pi_{\text{LLM}}(F^\sharp \mid p) \sim \Pr\left[F^\sharp \middle| p,\ \theta_{\text{LLM}},\ \mathcal{D}_{\text{train}},\ \pi_{\text{decode}},\ \epsilon,\ \psi\right].$$

Here, $\{F^\sharp_{cand}\} \subseteq \mathcal{H}$, which is the generated set of candidates. $\theta_{\text{LLM}}$ denotes the model parameters learned from its pretraining corpus $\mathcal{D}_{\text{train}}$, which capture the model's basic knowledge of language and semantics. $\pi_{\text{decode}}$ represents the internal decoding policy (e.g., temperature, nucleus sampling, beam width), and $\epsilon$ models the stochastic factors introduced during token generation. The term $\psi$ denotes outer feedback that implicitly modifies the reward function, such as reinforcement learning from human feedback (RLHF), instruction tuning, or system-level prompting, which biases the model toward human-aligned or task-specific behaviors. Together, these factors jointly determine the conditional generation distribution $\Pi_{\text{LLM}}(F^\sharp \mid p)$, explaining various performance (candidate sets with varying correctness and soundness rate) of different models under identical prompting conditions. In other word, $\Pi_{\text{LLM}}$ induces a specific probability distribution over the search space $\mathcal{H}$ of candidate transformers.

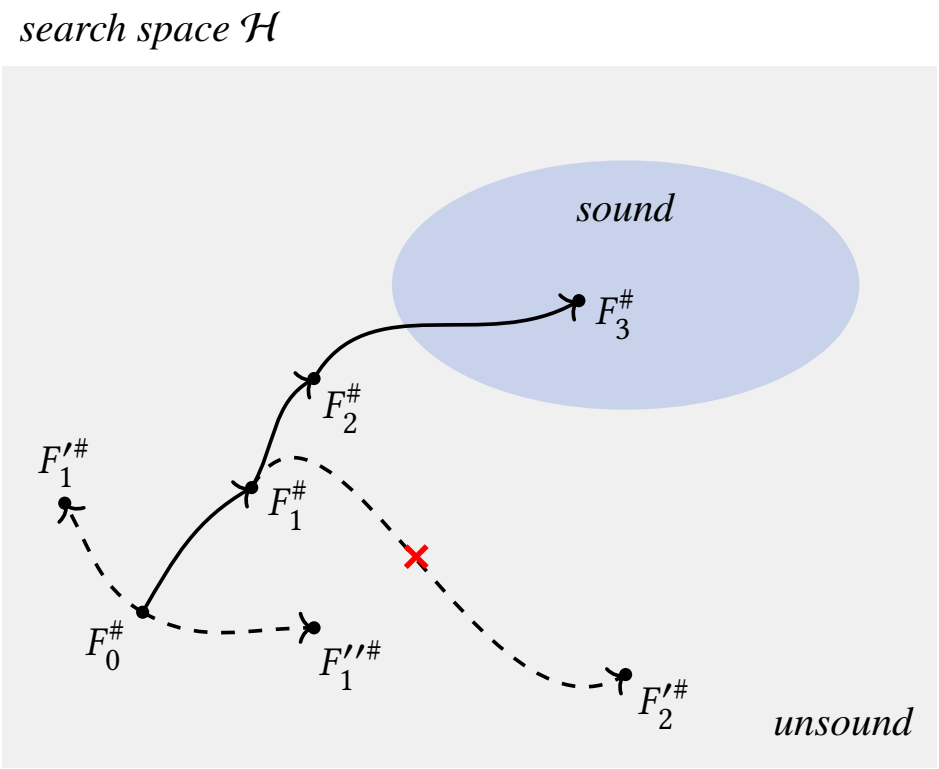

Fig. 5. Search trajectory of abstract transformers within the search space $\mathcal{H}$. Starting from $F^\sharp_0$, the process iteratively selects the candidate transformer with the lowest cost function value each round. Each refinement step must satisfy the progress condition $\mathcal{L}(\mathcal{R}(F^\sharp)) < \mathcal{L}(F^\sharp) - \lambda$; for example, the transition from $F^\sharp_1$ to $F'^\sharp_2$ is invalid since it violates this constraint.

*Cost Function.* As described in §4.2, we define a domain-specific cost function: $\mathcal{L}(F^\sharp) = \Delta_S(F^\sharp)$, where $\Delta_S(F^\sharp)$ measures the soundness deviation of the candidate transformer. which captures how far $F^\sharp$ deviates from the soundness condition, and is zero if and only if $F^\sharp$ is sound.

*Optimization Objective.* The synthesis problem is formulated as minimizing the soundness loss:

$$F^{\sharp *} = \arg\min_{F^\sharp} \mathcal{L}(F^\sharp) \quad \text{s.t. } F^\sharp \text{ satisfies all syntactic and semantic validity constraints.}$$

Here $F^{\sharp *}$ denotes the sound transformer. For any $F^\sharp \in \mathcal{U}$, the positive $\Delta_S(F^\sharp)$ quantifies the degree of unsoundness, guiding the refinement process to iteratively reduce $\Delta_S$ until the transformer enters $\mathcal{G}$.

*Search Strategy.* The cost function $\mathcal{L}$ is generally non-differentiable, so technically any gradient search strategy can not be utilized to explore the space of candidate transformers. However, to guarantee convergence while preserving exploration flexibility, we design a refinement strategy. Formally, given an initial candidate $F^\sharp_0 \in \mathcal{H}$, the synthesis system iteratively refines the transformer using a refinement operator $\mathcal{R} : \mathcal{H} \to \mathcal{H}$, which aims to monotonically decrease the loss:

$\mathcal{L}(\mathcal{R}(F^\sharp)) < \mathcal{L}(F^\sharp)$. To ensure the synthesis terminates within a finite number of steps, we require each refinement to achieve a minimum improvement on the cost function:

$$\mathcal{L}(\mathcal{R}(F^\sharp)) < \mathcal{L}(F^\sharp) - \lambda, \tag{5}$$

where $\lambda > 0$ is a fixed threshold controlling the minimum progress per iteration. The search trajectory is visualized in Fig. 5. Each step applies $\mathcal{R}$ to improve the transformer. When progress stalls (i.e., no decrease for $K$ consecutive rounds), the system can either terminate or adaptively reduce $\lambda$ to explore alternative refinement paths. The process continues until a sound transformer is found (i.e., $F_k^\sharp \in \mathcal{G}$) or a maximum number of attempts is reached. We therefore proceed to prove the convergence of the algorithm under this setting.

## 4.4 Convergence Proof

*Requirements.* The convergence proof relies on the following requirements:

(R1) For all constraint terms $C_{ij}$ and evaluated components $f(\mathbf{x})_i$ in Equation 4, both values are finite, i.e., $f(\mathbf{x})_i, C_{ij} \in \mathbb{R}$.

(R2) The aggregation operators $\bigoplus$ are all non-negative and bounded, i.e., $\forall S, 0 < \bigoplus S < \infty$, and the aggregated deviation is finite and non-negative, i.e., $\forall S_1, S_2, S_1 \subseteq S_2 \implies \bigoplus S_1 \subseteq \bigoplus S_2$. $\bigoplus S = 0$ if and only if $S = \emptyset$.

Theorem 4.1. *Assume each refinement step $\mathcal{R}$ satisfies the rule $\mathcal{L}(\mathcal{R}(F^\sharp)) < \mathcal{L}(F^\sharp) - \lambda$, where $\forall F^\sharp, 0 < \mathcal{L}(F^\sharp) < \infty$ and $\lambda > 0$. With fixed $\lambda$, the refinement process reaches $\mathcal{L}(F_T^\sharp) = 0$ in at most $T \leq \left\lceil \frac{\mathcal{L}(F_0^\sharp)}{\lambda} \right\rceil$ successful refinement steps. In particular, $\mathcal{L}(F_T^\sharp) = 0$ and hence $F_T^\sharp \in \mathcal{G}$.*

Proof. Let $L_t := \mathcal{L}(F_t^\sharp)$ denote the cost at the $t$-th successful refinement. By the improvement rule (R3), for every such step we have $\mathcal{L}(F_{t+1}^\sharp) < \mathcal{L}(F_t^\sharp) - \lambda$. Unrolling the inequality yields $\mathcal{L}(F_t^\sharp) < \mathcal{L}(F_0^\sharp) - t\lambda$. Choose $T := \left\lceil \frac{\mathcal{L}(F_0^\sharp)}{\lambda} \right\rceil$. Then $\mathcal{L}(F_T^\sharp) < \mathcal{L}(F_0^\sharp) - T\lambda \leq 0$. Combined with non-negativity (R2), we obtain $\mathcal{L}(F_T^\sharp) = 0$. By the definition of the cost (zero iff sound), this implies $F_T^\sharp$ is sound, i.e., $F_T^\sharp \in \mathcal{G}$. □

The parameter $\lambda$ controls the trade-off between convergence speed and refinement performance. A larger $\lambda$ enforces a stronger improvement per refinement step, leading to faster convergence in theory but making it harder for refinement to satisfy the required reduction. In contrast, a smaller $\lambda$ allows more gradual progress and increases the likelihood of finding valid refinements, though at the cost of slower convergence and more iterations. In our implementation based on LLMs, $\lambda$ is fixed. The model performs refinement within a fixed maximum number of attempts; if convergence is not achieved after reaching this limit, the process terminates.

## 4.5 Instantiation

*Relaxation Strategy.* In practice, directly evaluating the loss function Equation 4 is intractable, since both the set of abstract elements $A^*$ and the concretization $\gamma(z)$ may be infinite. To make computation feasible, we introduce a relaxation strategy by sampling from a finite subset of concretizations, denoted $\gamma_{\text{sample}}(z) \subseteq \gamma(z)$. For each sampled point $x_i$, where $\mathbf{x} \in \gamma_{\text{sample}}(z)$, we assign a normalized weight $w(x_i)$ that reflects its relative contribution to unsoundness. Intuitively, the weight function highlights those inputs that are more semantically critical, such as values close to nonlinear activation thresholds or regions prone to errors. Therefore the relaxed loss is:

$$\widetilde{\Delta_S}(F^\sharp) = {}^{z}\bigoplus_{z\in A^*} {}^{\mathbf{x}}\bigoplus_{\mathbf{x}\in\gamma_{sample}(z)} {}^{j}\bigoplus_{j=1}^{k} w(x_i)\cdot\varepsilon\big(f(\mathbf{x})_i, C_{ij}\big). \tag{6}$$

This relaxation ensures a tractable loss, while the weighting strategy preserves the informativeness of the deviation measure by emphasizing sampled points that contribute most to potential violations.

*Weight Function.* The weights are normalized as $w(x_i) = \frac{\phi(f,x_i)}{\sum_{x'\in\gamma_{\text{sample}}(z)}\phi(f,x_i')}$. Here, $f$ denotes the concrete function under analysis, $\phi(f, x_i)$ is a function-specific score designed to prioritize semantically critical configurations—such as those that cross nonlinear boundaries (e.g., zero for ReLU function), introduce branching behavior, or expose unstable symbolic patterns. In our formulation, we define $\phi(f, x_i)$ based on the sensitivity of the operator with respect to its input, measured by the numerical gradient: $\frac{\partial f(x)}{\partial x_i} \approx \frac{f(x_i+\epsilon)-f(x_i-\epsilon)}{2\epsilon}$. To ensure robustness, we apply the softplus transformation to the gradient magnitude, yielding: $\phi(f, x_i) = \log\big(1 + \exp(\|\nabla_{x_i} f(x_i)\|)\big)$. This construction leverages gradient information to emphasize inputs where the operator is most sensitive, while the softplus function prevents vanishing contributions in flat regions (where the gradient would otherwise be zero). As a result, the weight function captures both local sensitivity and stability, assigning higher importance to configurations likely to expose unsoundness in $F^\sharp$.

Algorithm 1 summarizes the whole constrained optimization procedure. In each round, we sample a set of candidate DSL transformers from the LLM, run validation to fix syntactic or semantic errors, discard those that cannot be repaired within the allowed number of attempts, and then invoke the soundness verifier. If a candidate is sound, we return it immediately and terminate; otherwise we score the unsound ones with the cost function, keep the best unsound candidate, and augment the next prompt with counterexamples, and repeat up to the retry budget. If no sound transformer is found, we return the default fallback. We set $\lambda = 0.0001$ in practice.

## 5 Evaluation

We evaluate our approach to answer the following research questions:

**RQ1**: How effective is SAIL procedure in guiding the synthesis process toward sound abstract transformers with different LLMs? (§5.1)

**RQ2**: Can SAIL generate abstract transformers for complex non-linear operations? (§5.2)

**RQ3**: How precise are the synthesized transformers for verifying neural networks? (§5.3)

**RQ4**: How does the performance change when the cost function or the validation module is ablated, and can LLMs still synthesize sound operators? (§5.4)

**RQ5**: How does SAIL compare to manual transformer design in terms of synthesis effort, and how well does it scale?

*Key Observations.* For RQ1, we consider the popular HardSigmoid activation for which no prior globally sound transformer exists. For RQ2, we evaluate on concrete operations, GeLU and ELU, which do not have any globally sound abstract transformer in the existing literature. For RQ3-4, we also consider additional operations for which handcrafted transformers exist. We find that as model capability increases, the dependence on cost-function feedback decreases overall (RQ1) but re-emerges prominently for complex operators (RQ2), highlighting that the cost function improves model capability overall through a formal process that provides explicit correctness guarantees. RQ3 demonstrates that SAIL achieves consistently high precision across a wide range of operators and abstract domains. For cases where handcrafted transformers exist, SAIL synthesized transformers match the precision of existing transformers. Our observations for RQ4 show that cost-function guidance with validation is essential for SAIL to synthesize diverse and sound abstract transformers. For RQ5, we find that SAIL significantly reduces manual effort with moderate synthesis overhead, while verification scales efficiently with model size.

**Algorithm 1** Multi-Round Abstract Transformer Generation and Validation

**Input:** Prompting template $P$; model client $M$; validator $Va$; soundness verifier $Ve$; cost evaluator $E$; minimum decrease $\lambda$; max retries $R$, fallback $F$.
**Output:** *result* (bool), *code* (DSL).

```
result ← false; code ← ∅; best_code ← ∅; best_score ← ∞
for r = 1 → R do
    Augment P with previous failures and counterexamples (if any)   ▷ use failed code to guide next round
    Generate completions {c_1, c_2, ...} from M using P
    for each completion c_i do
        Extract candidate DSL code d from c_i
        if d is empty then
            continue                                                  ▷ skip empty extraction
        for r = 1 → R do
            isvalid, d ← Va(d)   ▷ check the syntax and semantic correctness, fix errors if invalid
        if ¬ isvalid then
            continue                                                  ▷ skip invalid ones
        (passed, ce) ← Ve(d)                                          ▷ verify soundness
        if passed then
            return (true, d)                                          ▷ sound transformer found
        else if ce ≠ ∅ then
            score ← E(d)                      ▷ evaluate unsound transformer with cost function
            if score < best_score − λ then
                best_score ← score; best_code ← d
                Save (d, ce) for next prompt augmentation       ▷ keep "best" unsound candidate
return (false, F)                             ▷ terminate and return the default fallback
```

*Experimental Setup.* Sail is implemented in Python and integrated with the ConstraintFlow verification engine. All experiments are conducted on a GPU cluster node equipped with four NVIDIA A100 GPUs (40 GB each), an AMD EPYC 7763 CPU, and 256 GB of RAM, running CUDA 12.2. We evaluate Sail on a suite of challenging and popular neural network activation functions, including challenging piecewise-linear activations and non-linear activations, for which relatively less work exists on designing globally sound abstract transformers, such as HardTanh, HardSigmoid, HardSwish, Gelu, and Elu. We choose four state-of-the-art models: GPT-5, GPT-4o, Llama4-Maverick, and Claude-Opus-4. Since ConstraintFlow cannot directly verify nonlinear activation functions, Gelu and Elu, due to relying on Z3 as the underlying SMT solver, we manually verify the soundness and provide counterexamples for these. We use stochastic decoding during generation to ensure fair and stable evaluation, and we repeat each synthesis task multiple times under identical configurations and report the best performance across runs.

### 5.1 Effectiveness of Cost-Function Guidance

We demonstrate the effectiveness of the proposed cost-function guidance using the synthesis of the HardSigmoid operator, a nontrivial piecewise-linear activation that had no prior hand-crafted transformer in the DeepPoly domain. Four models (GPT-5, GPT-4o, Llama4-Maverick, and Claude-Opus-4) were tasked under identical configurations. Fig. 6 visualizes their synthesis

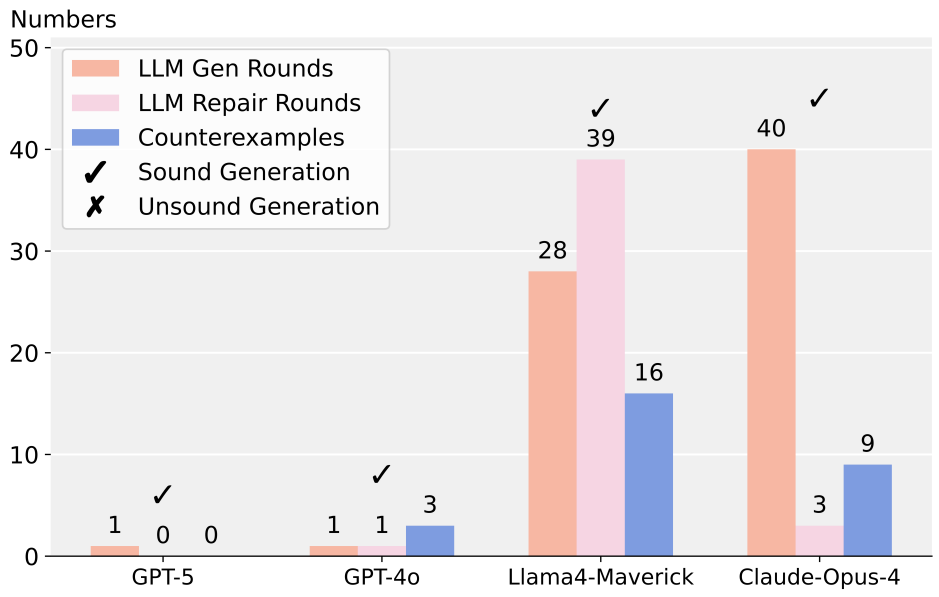

(a) Overall synthesis statistics across models.

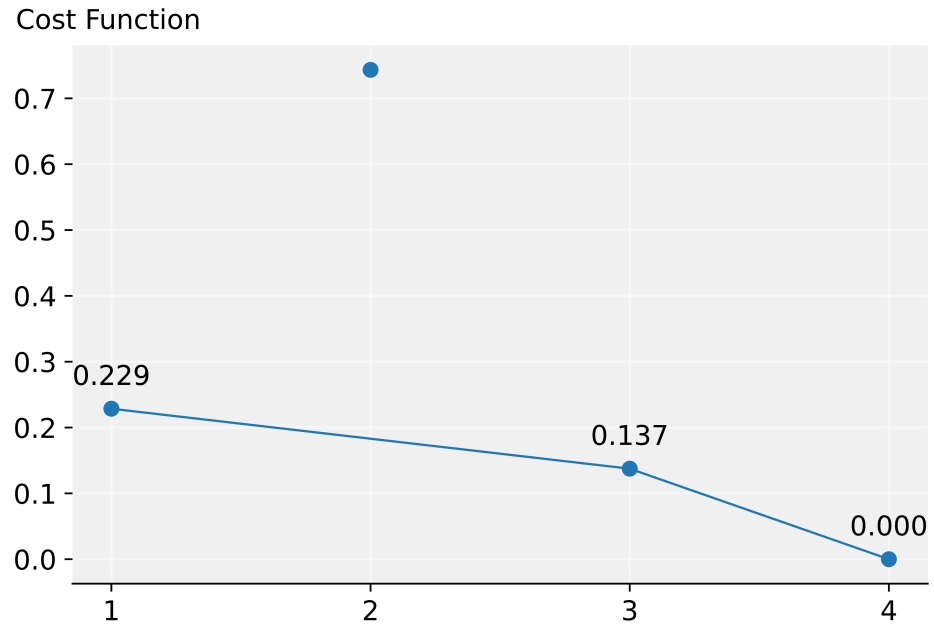

(b) Cost-function trajectory for GPT-4o.

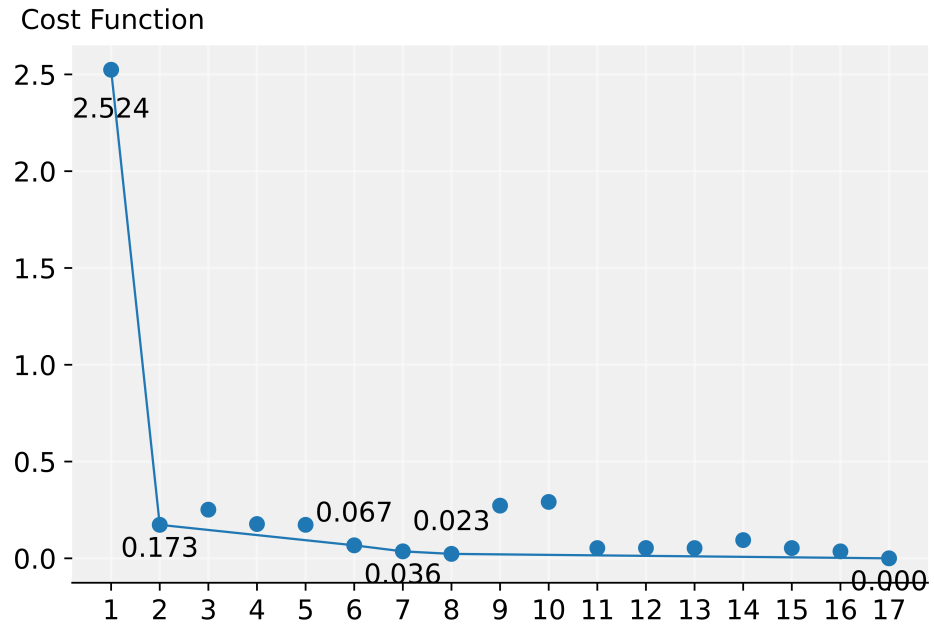

(c) Cost-function trajectory for Llama4-Maverick.

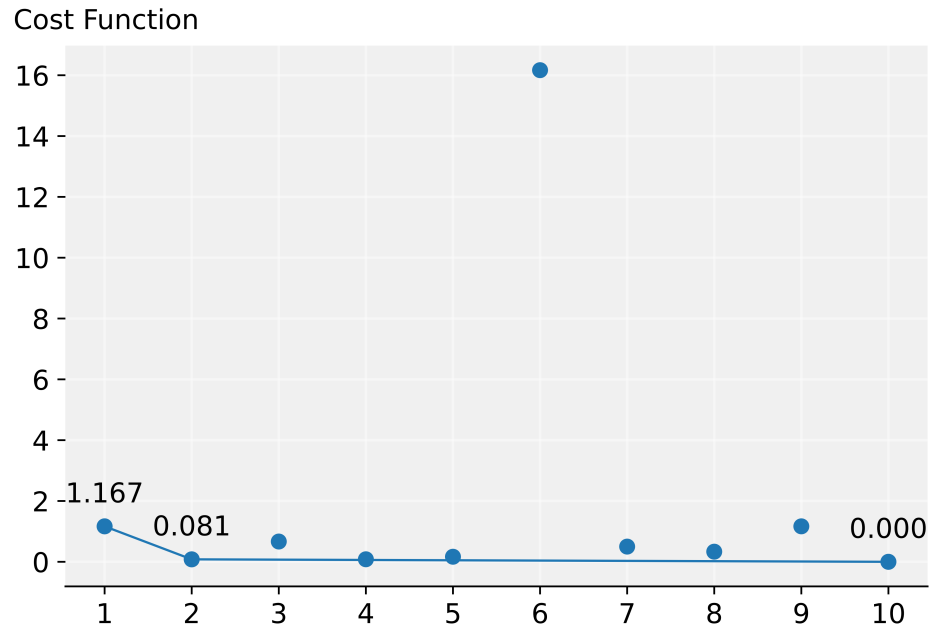

(d) Cost-function trajectory for Claude-Opus-4.

Fig. 6. Synthesis performance of different LLMs on HardSigmoid operators in the DeepPoly domain.

behavior, where Fig. 6a reports the number of generation, repair, and counterexamples rounds, and Fig. 6b, Fig. 6c, Fig. 6d show the cost trajectories for representative runs.

Across all models, GPT-5 completes synthesis without explicit help of counterexamples, while GPT-4o converges within several rounds with some feedback. In contrast, Llama4-Maverick and Claude-Opus-4 require substantially more verification feedback and repair rounds, exhibiting multiple cost drops triggered by counterexamples. These results reveal that model capability strongly influences the reliance on formal guidance. By following the optimization rule defined in Equation 5, the synthesis process discards occasional abnormal evaluations arising from sampling randomness, thereby maintaining a consistently decreasing and convergent cost trajectory. For example, as shown in Fig. 6d, the cost-function trajectory for Claude-Opus-4 exhibits a stable downward trend after filtering out such outliers. Overall, the results demonstrate that while stronger models such as GPT-5 can generate sound transformers autonomously, cost-function feedback remains essential for less capable models. The cost function thus compensates for model limitations and ensures convergence in complex synthesis tasks. Comprehensive results for the synthesis of other sound abstract interpreters across multiple models and domains can be found in Appendix J.

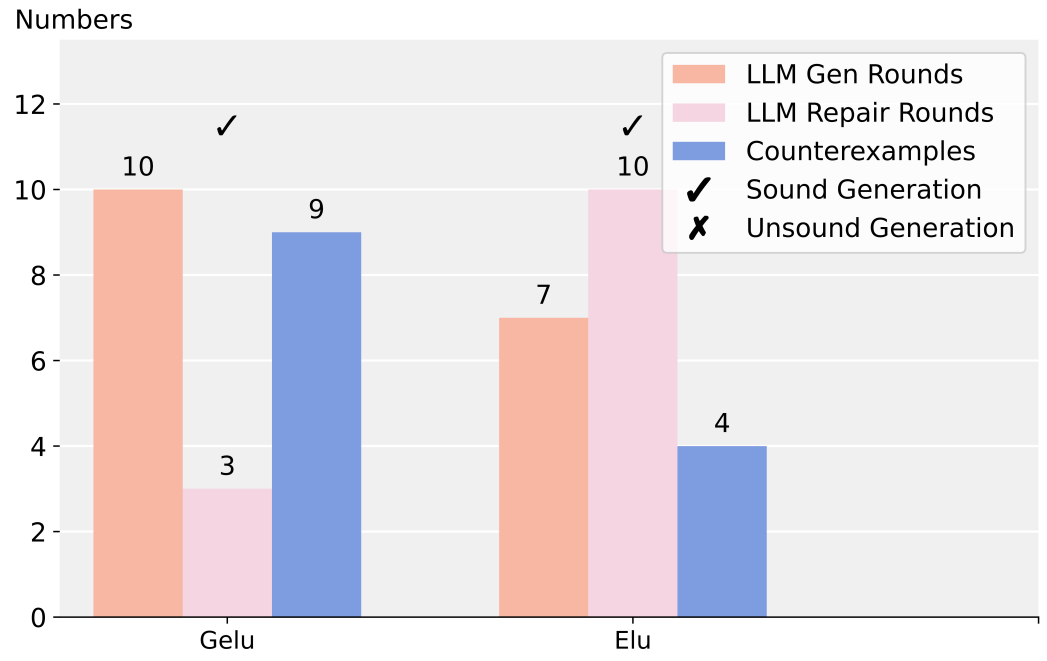

Fig. 7. Performance of GPT-5 generating DeepPoly transformers for GELU and ELU.

## 5.2 Handling Novel Nonlinear Operators

To further test the limits of SAIL 's synthesis capability for handling complex nonlinear operations, we task the best-performing model, GPT-5, with generating DeepPoly transformers for two representative nonlinear operators: GeLU and ELU. Formally, the GeLU function is defined as $\mathrm{GELU}(x) = \frac{1}{2}x(1 + \mathrm{erf}(x/\sqrt{2}))$, where erf denotes the Gaussian error function. The ELU function is defined as $\mathrm{ELU}(x) = x$ for $x > 0$ and $\alpha(e^x - 1)$ for $x \leq 0$ with the default $\alpha = 1$. The synthesis performance is shown in Fig. 7. The synthesis result for GeLU is shown in Fig. 8. And the synthesis result for ELU is shown in Fig. 10 in Appendix F. While GPT-5 can generate sound piecewise-linear transformers without explicit cost-function guidance (as shown in Fig. 6a), as the complexity of the target operator increases, the reliance on cost-function feedback re-emerges and increases as the complexity of the target operator grows. This trend highlights that while GPT-5 exhibits strong structural understanding, the cost-function feedback is essential for soundly capturing the behavior of novel and complicated nonlinear operators such as GeLU.

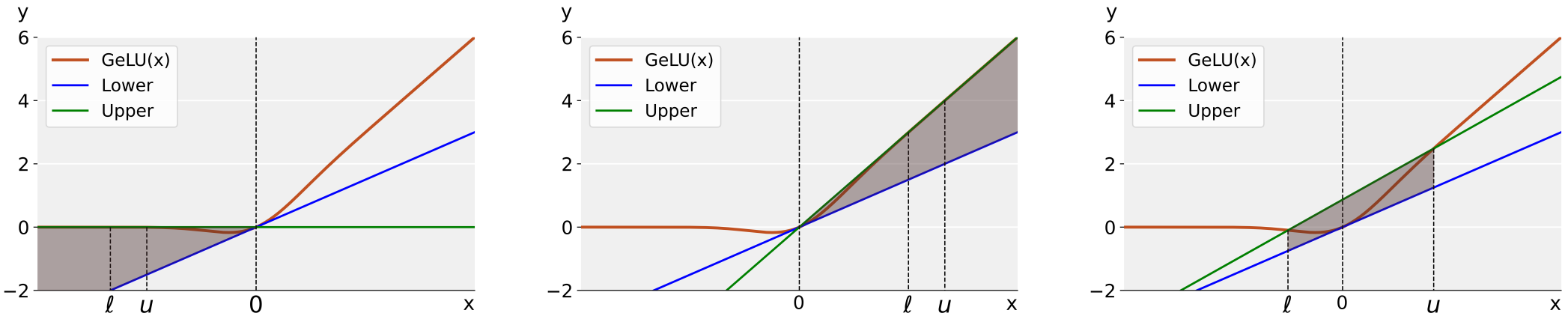

(a) $l < u < 0$. (b) $0 < l < u$. (c) $l < 0 < u$.

Fig. 8. **DeepPoly Transformer for GeLU.** The transformer considers three cases: (a) for $l < u < 0$, lower bound and upper bound are $y = 0.5x$ and $y = 0$ respectively; (b) for $0 < l < u$, lower bound and upper bound are $y = 0.5x$ and $y = x$ respectively; (c) for the mixed case ($l < 0 < u$), the upper bound is the secant line connecting $(l, \mathrm{GeLU}(l))$ and $(u, \mathrm{GeLU}(u))$, while the lower bound remains $y = 0.5x$. Since $\mathrm{GeLU}(x)$ is monotonic, the interval bounds correspond directly to $\mathrm{GeLU}(l)$ and $\mathrm{GeLU}(u)$. Each shaded region shows the area enclosed by the DeepPoly's polyhedra bounds, visualizing how they approximate the GeLU activation.

## 5.3 Evaluating Precision for Verifying Neural Networks

There exists no "best" transformer, therefore to assess the performance of synthesized transformers quantitatively, we measure the precision of GPT-5-generated transformers under the DeepPoly domain for verifying neural networks. We compare against handcrafted transformers provided by CONSTRAINTFLOW, which offers globally sound and the most precise transformers. More importantly, we show the precision of generated transformers for non-linear and complicated operators, whose corresponding abstract transformers do not exist in the literature, demonstrating the efficiency of SAIL. The underlying verification problem is standard image robustness verification: for a correctly classified input image, we check whether the network predicts the correct label for all perturbed inputs within a perturbation ball of radius $\epsilon \in \mathbb{R}$ around the original input. In our evaluation, precision is defined as the fraction of baseline-correct test inputs (i.e., samples correctly classified by the original network without the perturbation) that can be certified under a given perturbation bound.

We evaluate all transformers across multiple network architectures and training regimes, typically used for measuring verification performance in the literature [9, 61] including both fully connected (FCN) and convolutional (Conv) networks trained on MNIST [18] and CIFAR10 [36] datasets. For MNIST, we apply the perturbation to the entire image, whereas for CIFAR10 we restrict the perturbation to a single pixel to reflect more localized robustness settings. We consider the full test set and set the batch size to 100. For perturbations, we use $\epsilon = 0.005$ for MNIST and $\epsilon = 8/255$ for CIFAR10, which are standard choices in robustness verification. Part of the results are summarized in Table 1, GPT-5-synthesized transformers have different syntactical forms but the same semantics as transformers provided by ConstraintFlow, achieving precision on par with handcrafted transformers for all existing operators. For novel non-linear operators without existing handcrafted transformers, GPT-5–synthesized transformers also achieve high precision, demonstrating Sail 's ability to synthesize sound transformers with good qualities beyond the scope of manually designed ones. Comprehensive results and an additional evaluation for transformers synthesized under the DeepZ domain can be found in Appendix D.

Table 1. Precision comparison across different Networks based on the DeepPoly domain.

| Dataset | Network | Training | Activation | Layers | Perturbation $\epsilon$ | Precision | |
|---|---|---|---|---|---|---|---|
| | | | | | | Our work | Handcrafted |
| MNIST | Conv | Standard | ReLU | 6 | 0.005 | 1.0000 | 1.0000 |
| | Conv | Standard | ReLU6 | 3 | 0.005 | 1.0000 | ✗ |
| | FCN_5×100 | DiffAI | HardTanh | 5 | 0.005 | 0.9500 | 0.9500 |
| | FCN_6×500 | PGD | HardSigmoid | 6 | 0.005 | 1.0000 | ✗ |
| | FCN_3×100 | Standard | GELU | 3 | 0.005 | 0.9400 | ✗ |
| | FCN_4×1024 | Standard | GELU | 4 | 0.005 | 1.0000 | ✗ |
| | FCN_3×100 | Standard | ELU | 3 | 0.005 | 0.1400 | ✗ |
| CIFAR10 | Conv | DiffAI | ReLU | 3 | 8/255 | 1.0000 | 1.0000 |
| | FCN_4×100 | Standard | ReLU6 | 4 | 8/255 | 0.4490 | ✗ |
| | FCN_7×1024 | Standard | HardTanh | 7 | 8/255 | 0.9231 | 0.9231 |
| | FCN_4×100 | Standard | HardSwish | 4 | 8/255 | 0.1154 | ✗ |
| | FCN_6×500 | PGD | HardSigmoid | 6 | 8/255 | 1.0000 | ✗ |
| | FCN_6×500 | PGD | GELU | 6 | 8/255 | 1.0000 | ✗ |
| | FCN_7×1024 | Standard | GELU | 7 | 8/255 | 0.9787 | ✗ |

### 5.4 Ablation Study

To assess the individual contributions of the cost function and the validation–repair mechanism, we conduct a three-way ablation study using the Llama4-Maverick model as the synthesis engine. We prompt the model to generate DeepPoly transformers for a set of representative operators under three settings: (1) *with cost function and validation–repair module*, representing our full system; (2) *without cost function but with validation–repair*, where the model relies solely on repair feedback to correct unsound generations; and (3) *without both cost function and validation–repair*, where the model depends purely on its intrinsic reasoning ability without any external feedback. The results are shown in Fig. 13, Fig. 14, and Fig. 15 in Appendix I, which exhibit a clear hierarchy across the three configurations. Comparing setting (1) with setting (2) demonstrates that the cost function substantially improves the soundness and consistency of the synthesized transformers. By quantitatively penalizing unsound behaviors and providing continuous optimization feedback, the cost function enables the model to refine candidates beyond mere syntactic validity, achieving soundness that generalizes across operators. In contrast, comparing setting (2) with setting (3) highlights the importance of the validation–repair mechanism: even without cost feedback, it ensures structural well-formedness in most cases and prevents the cascade of syntax and type errors that otherwise dominate unconstrained generation to some extent. However, due to the absence of feedback, the generated transformer remains unsound. Together, these results confirm that structured repair and cost-guided optimization address complementary aspects of synthesis.

### 5.5 Computational Cost and Scalability Analysis

We evaluate the synthesis overhead and the downstream verification cost to demonstrate the practical efficiency of SAIL. Table 4 in Appendix E reports the runtime of the best result obtained from GPT-5-based synthesis of common abstract transformers in the DeepPoly domain, as well as the average verification cost on CIFAR-10 dataset across different networks. It shows that each synthesis in SAIL is completed within only a few seconds to at most a few hours, depending on operator complexity. The peak memory usage throughout synthesis is approximately 650 MB, which is modest and negligible compared to the typical system memory. Independent of SAIL, the verification cost scales with the size of the network (e.g., number of neurons) and is influenced by the performance of the underlying verification engine CONSTRAINTFLOW, which currently executes on CPU. Overall, compared to domain experts spending weeks or even months deriving mathematical abstract transformers [42, 73, 78], proving soundness, and implementing them efficiently, SAIL automates this process within hours and outputs executable DSL transformers that can be directly executed, significantly reducing human effort and achieving state-of-the-art performance with modest computational overhead. Moreover, SAIL can easily generalize to diverse operators and abstract domains. Each synthesis is a one-time cost per operator, after which the synthesized sound transformer can be reused across arbitrary models and verification tasks without additional effort.

## 6 Beyond Neural Network Verification

While SAIL uses neural network verification as a benchmark, both the synthesized abstract interpreters and the underlying framework can naturally extend to a broader range of tasks. The synthesized transformers can be integrated within other types of verifiers, such as for output range analysis [21] or for analyzing neurosymbolic programs [75]. The underlying framework can be applied to synthesize sound abstract transformers for concrete functions in other applications, as the core cost function Equation 4 is derived in a principled manner from the formal definition of global soundness, independently of any downstream tasks. We provide one example below by considering abstract transformers for derivatives of activation functions in differentiable programs [6].

Automatic differentiation (AD) and differentiable programming are widely adopted in machine learning, computer graphics, and scientific computing [1, 40, 69]. Abstract interpretation provides a principled framework for soundly reasoning about AD and verifying program properties expressed over derivatives [37], such as sensitivity, monotonicity, etc. However, sound abstract interpreters that can efficiently compute higher-order derivatives are still lacking in the literature. We demonstrate the potential of SAIL in this setting by successfully synthesizing a globally sound abstract transformer based on the Interval domain for the first-order derivative of the Sigmoid function. Analyzing gradients is essential in differentiable programs when explaining model behavior or verifying gradient-dependent properties (e.g., Lipschitz bounds or robustness), We choose Sigmoid as it is commonly used, and its derivative involves a non-trivial compositional structure due to its nonlinear form. Formally, the formulation in SAIL remains unchanged, with the term $f(\mathbf{x})_i$ in the cost function Equation 4 being instantiated with $\sigma(x_i)\big(1-\sigma(x_i)\big)$, where $\sigma(x_i) = \frac{1}{1+e^{-x_i}}$ and $i$ indexes the input dimensions. Since the derivative is applied pointwise, we still focus on a single $i$-th dimension during synthesis. Empirically, we manually verify soundness and provide counterexamples, as the presence of exponentials in the Sigmoid function makes the verification problem undecidable in general, similar to the cases of GeLU and ELU. Since the Interval domain is relatively simple for stronger models, we use Llama4-Maverick in this experiment to better demonstrate the effectiveness of cost-based guidance. The experiment shows that SAIL converges in 25 rounds on average, yielding a sound and precise transformer with case distinctions, as shown in Fig. 11 in Appendix G, demonstrating the generality of SAIL.

## 7 Related Work

*Program and Transformer Synthesis.* Program synthesis aims to automatically construct programs that satisfy formal specifications. Syntax-guided synthesis (SyGuS) [2] integrates semantic constraints with syntactic templates defined by a user-supplied grammar, enabling solver-guided exploration of a constrained search space. The dominant framework, Counterexample-Guided Inductive Synthesis (CEGIS) [66], alternates between candidate generation and verification, refining hypotheses using counterexamples until convergence. Within abstract interpretation, these ideas are applied to automatically construct sound abstract transformers for given operators and domains. Amurth [34] formulates this process as DSL-constrained synthesis, combining inductive refinement with formal verification to derive the most precise sound transformer expressible in a given DSL. Amurth2 [35] generalizes this method to reduced product domains via dual CEGIS loops that coordinate synthesis of soundness and precision across components. However, both approaches carry the risk of not converging. Recent work, such as *USTAD* [24] and LinSyn [51] extend this direction toward numerical and neural domains. USTAD develops a differentiable parametric framework for synthesizing sound linear transformers over polyhedral domains, while LinSyn automates the synthesis of tight linear bounds for arbitrary neural activations using constraint solving and local optimization verified by SMT (dReal). Another framework [50] combines interval-based verification with model finetuning to learn region-specific linear bounds, but guarantees only local soundness. Together, these frameworks trace a progression from symbolic [43] to solver- and optimization-driven synthesis, aiming to balance formal soundness with scalability in constructing numerical transformers.

*LLMs for Formal Reasoning and Verification.* Large language models (LLMs) have recently been explored as assistants for formal reasoning and program verification. Frameworks such as Verifier-in-the-Loop [71], Self-Refine [44], and GPT-f [52] integrate automated verifiers or theorem provers with iterative generation, allowing models to receive structured feedback when an output violates constraints. Similarly, Refine4LLM [10] couples LLMs with symbolic solvers and proof checkers to synthesize or repair proofs, verification conditions, and formal specifications. These systems demonstrate that integrating reasoning modules into the LLM generation loop can substantially enhance factual consistency and logical soundness. Nonetheless, most of these approaches are confined to discrete symbolic reasoning. For example, generating proofs or verification conditions rather than reasoning about continuous or numerical abstractions. Another central difficulty remains mitigating hallucination [74], where models produce seemingly valid proofs or specifications that fail semantic verification. Recent efforts [44, 52, 71, 72] address this issue through verifier-guided refinement, static analysis, confidence-based rejection sampling, etc.

## 8 Conclusion

Overall, Sail casts transformer synthesis as a constrained optimization problem driven by a novel soundness-based cost function based on LLMs, enabling globally sound and complex new transformers that can be instantiated for any abstract domain. The underlying ideas under Sail extend beyond neural network certification to any setting requiring automated construction of sound algorithms.

### Acknowledgments

We would like to thank the anonymous reviewers for their constructive feedback. This work was supported by an Amazon Research Award, an award from the Amazon-Illinois Center on AI for Interactive Conversational Experiences (AICE), and NSF Grants No. CCF-2238079, CCF-2316233, NAIRR240476, and an Open Philanthropy research grant.

## Data-Availability Statement

The artifact, including source code, models, and evaluation scripts, is publicly available on Zenodo [27] (DOI: 10.5281/zenodo.19591287), with instructions to reproduce the results. The latest version of the framework is maintained on GitHub at https://github.com/uiuc-focal-lab/SAIL.

## References

[1] Martín Abadi, Paul Barham, Jianmin Chen, Zhifeng Chen, Andy Davis, Jeffrey Dean, Matthieu Devin, Sanjay Ghemawat, Geoffrey Irving, Michael Isard, Manjunath Kudlur, Josh Levenberg, Rajat Monga, Sherry Moore, Derek G. Murray, Benoit Steiner, Paul Tucker, Vijay Vasudevan, Pete Warden, Martin Wicke, Yuan Yu, and Xiaoqiang Zheng. 2016. TensorFlow: A system for large-scale machine learning. https://arxiv.org/abs/1605.08695

[2] Rajeev Alur, Rastislav Bodik, Garvit Juniwal, Milo M. K. Martin, Mukund Raghothaman, Sanjit A. Seshia, Rishabh Singh, Armando Solar-Lezama, Emina Torlak, and Abhishek Udupa. 2013. Syntax-guided synthesis. In *2013 Formal Methods in Computer-Aided Design*. 1–8. doi:10.1109/FMCAD.2013.6679385

[3] Rajeev Alur, Arjun Radhakrishna, and Abhishek Udupa. 2017. Scaling enumerative program synthesis via divide and conquer. In *International conference on tools and algorithms for the construction and analysis of systems*. 319–336.

[4] Jacob Austin, Augustus Odena, Maxwell Nye, Maarten Bosma, Henryk Michalewski, David Dohan, Ellen Jiang, Carrie Cai, Michael Terry, Quoc Le, et al. 2021. Program Synthesis with Large Language Models. https://arxiv.org/abs/2108.07732

[5] Clark Barrett, Swarat Chaudhuri, Fabrizio Montesi, Jim Grundy, Pushmeet Kohli, Leonardo de Moura, Alexandre Rademaker, and Sorrachai Yingchareonthawornchai. 2026. CSLib: The Lean Computer Science Library. https://arxiv.org/abs/2602.04846

[6] Atilim Gunes Baydin, Barak A. Pearlmutter, Alexey Andreyevich Radul, and Jeffrey Mark Siskind. 2018. Automatic differentiation in machine learning: a survey. https://arxiv.org/abs/1502.05767

[7] Bruno Blanchet, Patrick Cousot, Radhia Cousot, Jérôme Feret, Laurent Mauborgne, Antoine Miné, David Monniaux, and Xavier Rival. 2002. *Design and implementation of a special-purpose static program analyzer for safety-critical real-time embedded software*. 85–108.

[8] Olivier Bouissou and Matthieu Martel. 2008. Abstract Interpretation of the Physical Inputs of Embedded Programs. In *Verification, Model Checking, and Abstract Interpretation*, Francesco Logozzo, Doron A. Peled, and Lenore D. Zuck (Eds.). 37–51.

[9] Christopher Brix, Stanley Bak, Taylor T. Johnson, and Haoze Wu. 2024. The Fifth International Verification of Neural Networks Competition (VNN-COMP 2024): Summary and Results. https://arxiv.org/abs/2412.19985

[10] Yufan Cai, Zhe Hou, David Sanan, Xiaokun Luan, Yun Lin, Jun Sun, and Jin Song Dong. 2025. Automated Program Refinement: Guide and Verify Code Large Language Model with Refinement Calculus. *Proc. ACM Program. Lang.* POPL, Article 69 (Jan. 2025). doi:10.1145/3704905

[11] Jose Cambronero, Hongyu Li, Seohyun Kim, Koushik Sen, and Satish Chandra. 2019. When Deep Learning Met Code Search. https://arxiv.org/abs/1905.03813

[12] Marco Campion, Mila Dalla Preda, and Roberto Giacobazzi. 2022. Partial (In)Completeness in abstract interpretation: limiting the imprecision in program analysis. *Proc. ACM Program. Lang.* POPL, Article 59 (Jan. 2022). doi:10.1145/3498721

[13] Patrick Cousot and Radhia Cousot. 1977. Abstract interpretation: a unified lattice model for static analysis of programs by construction or approximation of fixpoints. In *Proceedings of the 4th ACM SIGACT-SIGPLAN Symposium on Principles of Programming Languages (POPL '77)*. 238–252. doi:10.1145/512950.512973

[14] Patrick Cousot and Radhia Cousot. 1979. Systematic design of program analysis frameworks. In *Proceedings of the 6th ACM SIGACT-SIGPLAN Symposium on Principles of Programming Languages (POPL '79)*. 269–282. doi:10.1145/567752.567778

[15] Patrick Cousot, Radhia Cousot, Jerôme Feret, Laurent Mauborgne, Antoine Miné, David Monniaux, and Xavier Rival. 2005. The ASTREÉ analyzer. In *Proceedings of the 14th European Conference on Programming Languages and Systems (ESOP'05)*. 21–30. doi:10.1007/978-3-540-31987-0_3

[16] George B Dantzig. 1973. *Fourier-Motzkin elimination and its dual.* Technical Report.

[17] DeepSeek-AI, Daya Guo, et al. 2025. DeepSeek-R1: Incentivizing Reasoning Capability in LLMs via Reinforcement Learning. https://arxiv.org/abs/2501.12948

[18] Li Deng. 2012. The MNIST Database of Handwritten Digit Images for Machine Learning Research [Best of the Web]. *IEEE Signal Processing Magazine* 6 (2012), 141–142. doi:10.1109/MSP.2012.2211477

[19] Alessandra Di Pierro and Herbert Wiklicky. 2000. Measuring the precision of abstract interpretations. In *International Workshop on Logic-Based Program Synthesis and Transformation*. 147–164.

[20] Pranav Garg, Daniel Neider, Parthasarathy Madhusudan, and Dan Roth. 2016. Learning invariants using decision trees and implication counterexamples. *ACM Sigplan Notices* 1 (2016), 499–512.

[21] Timon Gehr, Matthew Mirman, Dana Drachsler-Cohen, Petar Tsankov, Swarat Chaudhuri, and Martin Vechev. 2018. AI2: Safety and Robustness Certification of Neural Networks with Abstract Interpretation. In *2018 IEEE Symposium on Security and Privacy (SP)*. 3–18. doi:10.1109/SP.2018.00058

[22] Robert Joseph George, Jennifer Cruden, Xiangru Zhong, Huan Zhang, and Anima Anandkumar. 2026. TorchLean: Formalizing Neural Networks in Lean. https://arxiv.org/abs/2602.22631

[23] Roberto Giacobazzi, Francesco Ranzato, and Francesca Scozzari. 2000. Making abstract interpretations complete. *J. ACM* 47, 2 (March 2000), 361–416. doi:10.1145/333979.333989

[24] Shaurya Gomber, Debangshu Banerjee, and Gagandeep Singh. 2025. Universal Synthesis of Differentiably Tunable Numerical Abstract Transformers. https://arxiv.org/abs/2507.11827

[25] Sven Gowal, Krishnamurthy Dvijotham, Robert Stanforth, Rudy Bunel, Chongli Qin, Jonathan Uesato, Relja Arandjelovic, Timothy Mann, and Pushmeet Kohli. 2019. On the Effectiveness of Interval Bound Propagation for Training Verifiably Robust Models. https://arxiv.org/abs/1810.12715

[26] Arya Grayeli, Atharva Sehgal, Omar Costilla-Reyes, Miles Cranmer, and Swarat Chaudhuri. 2024. Symbolic Regression with a Learned Concept Library. https://arxiv.org/abs/2409.09359

[27] Qiuhan Gu, Avaljot Singh, and Gagandeep Singh. 2026. SAIL: Sound Abstract Interpreters with LLMs. doi:10.5281/zenodo.19591287

[28] Sumit Gulwani, Oleksandr Polozov, Rishabh Singh, et al. 2017. Program synthesis. *Foundations and Trends® in Programming Languages* 4, 1-2 (2017), 1–119.

[29] Kihong Heo, Hakjoo Oh, and Kwangkeun Yi. 2017. Machine-Learning-Guided Selectively Unsound Static Analysis. In *2017 IEEE/ACM 39th International Conference on Software Engineering (ICSE)*. 519–529. doi:10.1109/ICSE.2017.54

[30] Gérard Huet, Gilles Kahn, and Christine Paulin-Mohring. 2002. The Coq proof assistant: a tutorial: version 7.2. (2002).

[31] Naman Jain, Skanda Vaidyanath, Arun Iyer, Nagarajan Natarajan, Suresh Parthasarathy, Sriram Rajamani, and Rahul Sharma. 2022. Jigsaw: Large language models meet program synthesis. In *Proceedings of the 44th International Conference on Software Engineering*. 1219–1231.

[32] Jacques-Henri Jourdan, Vincent Laporte, Sandrine Blazy, Xavier Leroy, and David Pichardie. 2015. A Formally-Verified C Static Analyzer. In *Proceedings of the 42nd Annual ACM SIGPLAN-SIGACT Symposium on Principles of Programming Languages (POPL '15)*. 247–259. doi:10.1145/2676726.2676966

[33] Julien Julien Bertrane, Patrick Cousot, Radhia Cousot, Jérôme Feret, Laurent Mauborgne, Antoine Miné, and Xavier Rival. 2011. Static analysis by abstract interpretation of embedded critical software. *SIGSOFT Softw. Eng. Notes* 1 (2011), 1–8. doi:10.1145/1921532.1921553

[34] Pankaj Kumar Kalita, Sujit Kumar Muduli, Loris D'Antoni, Thomas Reps, and Subhajit Roy. 2022. Synthesizing abstract transformers. *Proc. ACM Program. Lang.* OOPSLA2, Article 171 (Oct. 2022). doi:10.1145/3563334

[35] Pankaj Kumar Kalita, Thomas Reps, and Subhajit Roy. 2025. Automated Abstract Transformer Synthesis for Reduced Product Domains. *ACM Transactions on Software Engineering and Methodology* (2025).

[36] Alex Krizhevsky and Geoffrey Hinton. 2009. *The CIFAR-10 and CIFAR-100 datasets*. https://www.cs.toronto.edu/~kriz/cifar.html

[37] Jacob Laurel, Rem Yang, Shubham Ugare, Robert Nagel, Gagandeep Singh, and Sasa Misailovic. 2022. A general construction for abstract interpretation of higher-order automatic differentiation. *Proc. ACM Program. Lang.* 6, OOPSLA2, Article 161 (Oct. 2022). doi:10.1145/3563324

[38] Joel Lehman, Jonathan Gordon, Shawn Jain, Kamal Ndousse, Cathy Yeh, and Kenneth O. Stanley. 2022. Evolution through Large Models. https://arxiv.org/abs/2206.08896

[39] Maikel Leon. 2025. GPT-5 and open-weight large language models: Advances in reasoning, transparency, and control. *Information Systems* (2025), 102620.

[40] Tzu-Mao Li, Michaël Gharbi, Andrew Adams, Frédo Durand, and Jonathan Ragan-Kelley. 2018. Differentiable programming for image processing and deep learning in halide. *ACM Trans. Graph.* 37, 4, Article 139 (July 2018). doi:10.1145/3197517.3201383

[41] Yujia Li, David Choi, Junyoung Chung, Nate Kushman, Julian Schrittwieser, Rémi Leblond, Tom Eccles, James Keeling, Felix Gimeno, Agustin Dal Lago, et al. 2022. Competition-level code generation with alphacode. *Science* 378, 6624 (2022), 1092–1097.

[42] Changliu Liu, Tomer Arnon, Christopher Lazarus, Christopher Strong, Clark Barrett, and Mykel J. Kochenderfer. 2020. Algorithms for Verifying Deep Neural Networks. https://arxiv.org/abs/1903.06758

[43] Sirui Lu and Rastislav Bodík. 2023. Grisette: Symbolic Compilation as a Functional Programming Library. *Proc. ACM Program. Lang.* 7, POPL, Article 16 (Jan. 2023). doi:10.1145/3571209

[44] Aman Madaan, Niket Tandon, Prakhar Gupta, Skyler Hallinan, Luyu Gao, Sarah Wiegreffe, Uri Alon, Nouha Dziri, Shrimai Prabhumoye, Yiming Yang, Shashank Gupta, Bodhisattwa Prasad Majumder, Katherine Hermann, Sean Welleck, Amir Yazdanbakhsh, and Peter Clark. 2023. Self-Refine: Iterative Refinement with Self-Feedback. https://arxiv.org/abs/2303.17651

[45] Antoine Miné. 2006. Symbolic methods to enhance the precision of numerical abstract domains. In *International Workshop on Verification, Model Checking, and Abstract Interpretation*. 348–363.

[46] Antoine Miné. 2017. Static Analysis of Embedded Real-Time Concurrent Software with Dynamic Priorities. *Electronic Notes in Theoretical Computer Science* 331 (2017), 3–39. doi:10.1016/j.entcs.2017.02.002 Proceedings of the Sixth Workshop on Numerical and Symbolic Abstract Domains (NSAD 2016).

[47] Matthew Mirman, Timon Gehr, and Martin T. Vechev. 2018. Differentiable Abstract Interpretation for Provably Robust Neural Networks. In *International Conference on Machine Learning*. https://api.semanticscholar.org/CorpusID:51872670

[48] Erik Nijkamp, Bo Pang, Hiroaki Hayashi, Lifu Tu, Huan Wang, Yingbo Zhou, Silvio Savarese, and Caiming Xiong. 2023. CodeGen: An Open Large Language Model for Code with Multi-Turn Program Synthesis. https://arxiv.org/abs/2203.13474

[49] Alexander Novikov, Ngân Vũ, Marvin Eisenberger, Emilien Dupont, Po-Sen Huang, Adam Zsolt Wagner, Sergey Shirobokov, Borislav Kozlovskii, Francisco J. R. Ruiz, Abbas Mehrabian, M. Pawan Kumar, Abigail See, Swarat Chaudhuri, George Holland, Alex Davies, Sebastian Nowozin, Pushmeet Kohli, and Matej Balog. 2025. AlphaEvolve: A coding agent for scientific and algorithmic discovery. https://arxiv.org/abs/2506.13131

[50] Brandon Paulsen and Chao Wang. 2022. Example Guided Synthesis of Linear Approximations for Neural Network Verification. In *Computer Aided Verification: 34th International Conference, CAV 2022, Haifa, Israel, August 7–10, 2022, Proceedings, Part I*. 149–170. doi:10.1007/978-3-031-13185-1_8

[51] Brandon Paulsen and Chao Wang. 2022. LinSyn: Synthesizing Tight Linear Bounds for Arbitrary Neural Network Activation Functions. https://arxiv.org/abs/2201.13351

[52] Stanislas Polu and Ilya Sutskever. 2020. Generative Language Modeling for Automated Theorem Proving. https://arxiv.org/abs/2009.03393

[53] Thomas Reps and Aditya Thakur. 2016. Automating Abstract Interpretation. In *Proceedings of the 17th International Conference on Verification, Model Checking, and Abstract Interpretation - Volume 9583 (VMCAI 2016)*. 3–40. doi:10.1007/978-3-662-49122-5_1

[54] Bernardino Romera-Paredes, Mohammadamin Barekatain, Alexander Novikov, Matej Balog, M. Pawan Kumar, Emilien Dupont, Francisco J. R. Ruiz, Jordan S. Ellenberg, Pengming Wang, Omar Fawzi, Pushmeet Kohli, and Alhussein Fawzi. 2024. Mathematical discoveries from program search with large language models. *Nat.* 7995 (January 2024), 468–475. https://doi.org/10.1038/s41586-023-06924-6

[55] Parshin Shojaee, Kazem Meidani, Shashank Gupta, Amir Barati Farimani, and Chandan K Reddy. 2025. LLM-SR: Scientific Equation Discovery via Programming with Large Language Models. https://arxiv.org/abs/2404.18400

[56] Avaljot Singh, Yasmin Sarita, Charith Mendis, and Gagandeep Singh. 2024. ConstraintFlow: A Declarative DSL for Easy Development of DNN Certifiers. 407–424. doi:10.1007/978-3-031-74776-2_16

[57] Avaljot Singh, Yasmin Chandini Sarita, Charith Mendis, and Gagandeep Singh. 2025. Automated Verification of Soundness of DNN Certifiers. *Proc. ACM Program. Lang.* OOPSLA1, Article 144 (April 2025). doi:10.1145/3720509

[58] Avaljot Singh, Yamin Chandini Sarita, Aditya Mishra, Ishaan Goyal, Gagandeep Singh, and Charith Mendis. 2025. A Tensor-Based Compiler and a Runtime for Neuron-Level DNN Certifier Specifications. https://arxiv.org/abs/2507.20055

[59] Gagandeep Singh and Deepika Chawla. 2025. Position: Formal Methods are the Principled Foundation of Safe AI. In *ICML Workshop on Technical AI Governance (TAIG)*. https://openreview.net/forum?id=7V5CDSsjB7

[60] Gagandeep Singh, Timon Gehr, Matthew Mirman, Markus Püschel, and Martin Vechev. 2018. Fast and effective robustness certification. In *Proceedings of the 32nd International Conference on Neural Information Processing Systems (NIPS'18)*. 10825–10836.

[61] Gagandeep Singh, Timon Gehr, Markus Püschel, and Martin Vechev. 2019. An abstract domain for certifying neural networks. *Proc. ACM Program. Lang.* POPL, Article 41 (Jan. 2019). doi:10.1145/3290354

[62] Gagandeep Singh, Timon Gehr, Markus Püschel, and Martin T. Vechev. 2018. Boosting Robustness Certification of Neural Networks. In *International Conference on Learning Representations*. https://api.semanticscholar.org/CorpusID:196059499

[63] Gagandeep Singh, Jacob Laurel, Sasa Misailovic, Debangshu Banerjee, Avaljot Singh, Changming Xu, Shubham Ugare, and Huan Zhang. 2025. Safety and Trust in Artificial Intelligence with Abstract Interpretation. *Found. Trends Program. Lang.* 3–4 (June 2025), 250–408. doi:10.1561/2500000062

[64] Gagandeep Singh, Markus Püschel, and Martin Vechev. 2017. Fast polyhedra abstract domain. In *Proceedings of the 44th ACM SIGPLAN Symposium on Principles of Programming Languages*. 46–59.

[65] Armando Solar-Lezama. 2009. The sketching approach to program synthesis. In *Asian symposium on programming languages and systems*. 4–13.

[66] Armando Solar-Lezama. 2013. Program sketching. *Int. J. Softw. Tools Technol. Transf.* 5–6 (Oct. 2013), 475–495. doi:10.1007/s10009-012-0249-7

[67] Benno Stein, Bor-Yuh Evan Chang, and Manu Sridharan. 2021. Demanded abstract interpretation. In *Proceedings of the 42nd ACM SIGPLAN International Conference on Programming Language Design and Implementation (PLDI 2021)*. 282–295. doi:10.1145/3453483.3454044

[68] Gemini Team, Rohan Anil, et al. 2025. Gemini: A Family of Highly Capable Multimodal Models. https://arxiv.org/abs/2312.11805
[69] Andrea Walther. 2009. Getting Started with ADOL-C. In *Combinatorial Scientific Computing (Dagstuhl Seminar Proceedings (DagSemProc), Vol. 9061)*. 1–10. doi:10.4230/DagSemProc.09061.10
[70] Wenhua Wang, Yuqun Zhang, Yulei Sui, Yao Wan, Zhou Zhao, Jian Wu, Philip S. Yu, and Guandong Xu. 2022. Reinforcement-Learning-Guided Source Code Summarization Using Hierarchical Attention. *IEEE Transactions on Software Engineering* 48, 1 (2022), 102–119. doi:10.1109/TSE.2020.2979701
[71] Yixuan Wang, Chao Huang, Zhaoran Wang, Zhilu Wang, and Qi Zhu. 2021. Verification in the Loop: Correct-by-Construction Control Learning with Reach-avoid Guarantees. https://arxiv.org/abs/2106.03245
[72] Guannan Wei, Zhuo Zhang, and Caterina Urban. 2025. Hallucination-Resilient LLM-Driven Sound and Tunable Static Analysis: A Case of Higher-Order Control-Flow Analysis. In *Proceedings of the 1st ACM SIGPLAN International Workshop on Language Models and Programming Languages (LMPL '25)*. 6–11. doi:10.1145/3759425.3763378
[73] Kaidi Xu, Huan Zhang, Shiqi Wang, Yihan Wang, Suman Jana, Xue Lin, and Cho-Jui Hsieh. 2021. Fast and Complete: Enabling Complete Neural Network Verification with Rapid and Massively Parallel Incomplete Verifiers. https://arxiv.org/abs/2011.13824
[74] Ziwei Xu, Sanjay Jain, and Mohan Kankanhalli. 2025. Hallucination is Inevitable: An Innate Limitation of Large Language Models. https://arxiv.org/abs/2401.11817
[75] Chenxi Yang and Swarat Chaudhuri. 2022. Safe Neurosymbolic Learning with Differentiable Symbolic Execution. https://arxiv.org/abs/2203.07671
[76] Shunyu Yao, Jeffrey Zhao, Dian Yu, Nan Du, Izhak Shafran, Karthik Narasimhan, and Yuan Cao. 2023. ReAct: Synergizing Reasoning and Acting in Language Models. https://arxiv.org/abs/2210.03629
[77] Huan Zhang, Hongge Chen, Chaowei Xiao, Sven Gowal, Robert Stanforth, Bo Li, Duane Boning, and Cho-Jui Hsieh. 2019. Towards Stable and Efficient Training of Verifiably Robust Neural Networks. https://arxiv.org/abs/1906.06316
[78] Huan Zhang, Tsui-Wei Weng, Pin-Yu Chen, Cho-Jui Hsieh, and Luca Daniel. 2018. Efficient Neural Network Robustness Certification with General Activation Functions. https://arxiv.org/abs/1811.00866

## A Prompt Templates

In this section, we show the prompt used for transformer generation and transformer repair in Sail.

**Generation Prompting Template**

*General instructions.*
You are a formal methods expert working on neural network verification. Your task is to generate the *[certifier]* transformers for DNN operators. Generate the transformer in Constraintflow DSL.

*[Information about the domain specific language.]*
Here is the grammar of Constraintflow DSL:

```
expr_list : expr COMMA expr_list
        | expr ;
exprs: expr exprs
        | expr;
metadata: WEIGHT
        | BIAS
        | EQUATIONS
        | LAYER ;
expr: FALSE #false
        | TRUE #true
        | IntConst #int
        | FloatConst #float
        | VAR #varExp
        | EPSILON #epsilon
        | CURR #curr
        | PREV #prev
        | PREV_0 #prev_0
        | PREV_1 #prev_1
        | CURRLIST #curr_list
        | LPAREN expr RPAREN #parenExp
        | LSQR expr_list RSQR #exprarray
        | expr LSQR metadata RSQR #getMetadata
        | expr LSQR VAR RSQR #getElement
        | expr binop expr #binopExp
        | NOT expr #not
        | MINUS expr #neg
        | expr QUES expr COLON expr #cond
        | expr DOT TRAV LPAREN direction COMMA expr COMMA expr COMMA expr RPAREN
LBRACE expr RBRACE #traverse
        | argmax_op LPAREN expr COMMA expr RPAREN #argmaxOp
        | max_op LPAREN expr RPAREN #maxOpList
        | max_op LPAREN expr COMMA expr RPAREN #maxOp
        | list_op LPAREN expr RPAREN #listOp
        | expr DOT MAP LPAREN expr RPAREN #map
        | expr DOT MAPLIST LPAREN expr RPAREN #map_list
        | expr DOT DOTT LPAREN expr RPAREN #dot
        | expr DOT CONCAT LPAREN expr RPAREN #concat
        | LP LPAREN lp_op COMMA expr COMMA expr RPAREN #lp
        | VAR LPAREN expr_list RPAREN #funcCall
```

| VAR exprs #curry ;
trans_ret :
expr QUES trans_ret COLON trans_ret #condtrans
| LPAREN trans_ret RPAREN #parentrans
| expr_list #trans ;

*[Information about the certifier. Using DeepPoly as an example below:]*
DeepPoly certifier uses four kinds of bounds to approximate the operator: (Float l, Float u, PolyExp L, PolyExp U). They must follow the constraints that: curr[l] <= curr <= curr[u] & curr[L] <= curr <= curr[U]. ‘curr‘ here means the current neuron, ‘prev‘ means the inputs to the operator. When the operator takes multiple inputs, use ‘prev_0‘, ‘prev_1‘, ... to refer to each input. So every transformer in each case of the case analysis must return four values. Use any funstions below if needed instead of use arithmetic operators. Function you can use:
- func simplify_lower(Neuron n, Float coeff) = (coeff >= 0) ? (coeff * n[l]) : (coeff * n[u]);
- func simplify_upper(Neuron n, Float coeff) = (coeff >= 0) ? (coeff * n[u]) : (coeff * n[l]);
- func replace_lower(Neuron n, Float coeff) = (coeff >= 0) ? (coeff * n[L]) : (coeff * n[U]);
- func replace_upper(Neuron n, Float coeff) = (coeff >= 0) ? (coeff * n[U]) : (coeff * n[L]);
- func priority(Neuron n) = n[layer];
- func priority2(Neuron n, Float c) = -n[layer];
- func stop(Neuron n) = false;
- func stop_traverse(Neuron n, Float c) = false;
- func backsubs_lower(PolyExp e, Neuron n) = (e.traverse(backward, priority2, stop_traverse, replace_lower)e <= n).map(simplify_lower);
- func backsubs_upper(PolyExp e, Neuron n) = (e.traverse(backward, priority2, stop_traverse, replace_upper)e >= n).map(simplify_upper);
- func f(Neuron n1, Neuron n2) = n1[l] >= n2[u];
- func slope(Float x1, Float x2) = ((x1 * (x1 + 3))-(x2 * (x2 + 3))) / (6 * (x1-x2));
- func intercept(Float x1, Float x2) = x1 * ((x1 + 3) / 6) - (slope(x1, x2) * x1);
- func f(Neuron n1, Neuron n2) = n1[l] >= n2[u];
- func f1(Float x) = x < 3 ? x * ((x + 3) / 6) : x;
- func f2(Float x) = x * ((x + 3) / 6);
- func f3(Neuron n) = max(f2(n[l]), f2(n[u]));
- func compute_l(Neuron n1, Neuron n2) = min([n1[l]*n2[l], n1[l]*n2[u], n1[u]*n2[l], n1[u]*n2[u]]);
- func compute_u(Neuron n1, Neuron n2) = max([n1[l]*n2[l], n1[l]*n2[u], n1[u]*n2[l], n1[u]*n2[u]]);
- func avg(List<Float> xs) = sum(xs) / len(xs);
- func argmax(List<Neuron> ns, (Neuron, Neuron -> Bool) cmp) = [ n | n in ns, forall m in ns. cmp(n, m) ];
- func argmin(List<Neuron> ns, (Neuron, Neuron -> Bool) cmp) = [ n | n in ns, forall m in ns. cmp(n, m) ];

Don’t add comments to DSL.

*[Two-shot prompting. Using DeepPoly as an example below:]*
### Example: Abs operator
Input: Generate the transformer for ‘abs‘ operator
Output:
"""
def Shape as (Float l, Float u, PolyExp L, PolyExp U)
{[(curr[l]<=curr),(curr[u]>=curr),(curr[L]<=curr),(curr[U]>=curr)]};

transformer deeppoly Abs -> ((prev[l]) >= 0) ? ((prev[l]), (prev[u]), (prev), (prev)) : (((prev[u]) <= 0) ? (0-(prev[u]), 0-(prev[l]), 0-(prev), 0-(prev)) : (0, max(prev[u], 0-prev[l]), prev, prev*(prev[u]+prev[l])/(prev[u]-prev[l]) - (((2*prev[u])*prev[l])/(prev[u]-prev[l]))) );
"""

### Example: Affine operator
Input: Generate the transformer for ‘affine‘ operator
Output:
"""
def Shape as (Float l, Float u, PolyExp L, PolyExp U) {[(curr[l]<=curr),(curr[u]>=curr),(curr[L]<=curr),(curr[U]>=curr)]};
transformer deeppoly Affine -> (backsubs_lower(prev.dot(curr[weight]) + curr[bias], curr), backsubs_upper(prev.dot(curr[weight]) + curr[bias], curr), prev.dot(curr[weight]) + curr[bias], prev.dot(curr[weight]) + curr[bias]);
"""

*[Query.]*
### Now generate the transformer for ‘{api}‘ operator
Input: Generate the transformer for ‘{api}‘ operator
Output: """

**Repair Prompting Template**

You are a DSL repair assistant. Fix the following DSL code based on the error.
[ERROR]:
*[Error messages.]*

[CODE]:
*[Incorrect generation.]*

Return only the fixed DSL code.

## B Generation Examples

In this section, we show three different kinds of transformer candidates generated by Sail.

```
transformer DeepPoly {
    HardSigmoid ->
        ((prev[u]) <= ((0 - attr[beta]) / attr[alpha])) ?
            (0, 0, 0, 0) :
        ((prev[l]) >= ((1 - attr[beta]) / attr[alpha])) ?
            (1, 1, 1, 1) :
        ((prev[l]) >= ((0 - attr[beta]) / attr[alpha])) ?
            ((prev[u]) <= ((1 - attr[beta]) / attr[alpha])) ?
                ((attr[alpha] * prev[l] + attr[beta]),
                 (attr[alpha] * prev[u] + attr[beta]),
                 (attr[alpha] * prev + attr[beta]),
                 (attr[alpha] * prev + attr[beta])) :
                ((attr[alpha] * prev[l] + attr[beta]),
                 1,
                 (attr[alpha] * prev + 1 - (attr[alpha] * prev[u])),
                 1) :
```

```
        ((prev[u]) <= ((1 - attr[beta]) / attr[alpha])) ?
            (0,
             (attr[alpha] * prev[u] + attr[beta]),
             0,
             (attr[alpha] * prev - (attr[alpha] * prev[l]))) :
            (0, 1, 0, 1);
}
```

Listing 1. Invalid transformer (Syntax error. Undefined variable: attr, alpha, beta)

```
transformer deeppoly {
    HardSigmoid -> (prev[u] <= -3) ? (0, 0, 0, 0)
        : ((prev[l] >= 3) ? (1, 1, 1, 1)
        : ((prev[u] <= 3)
            ? ((prev[l] >= -3)
                ? ((prev[l] + 3) / 6, (prev[u] + 3) / 6, (prev + 3) / 6, (prev + 3) / 6)
                : (0, (prev[u] + 3) / 6, (prev + 3) / 6,
                   prev[l] * ((prev[u] + 3) / (6 * (prev[u] - prev[l]))) * prev
                   - ((prev[u] + 3) / (6 * (prev[u] - prev[l]))) * prev[l]))
            : ((prev[l] >= -3)
                ? ((prev[l] + 3) / 6, 1,
                   ((3 - prev[l]) / (6 * (prev[u] - prev[l]))) * (prev - prev[l]) + (prev[l] + 3) / 6,
                   (prev + 3) / 6)
                : (0, 1, (prev + 3) / (prev[u] + 3), (prev - prev[l]) / (3 - prev[l])))));
}
```

Listing 2. Unsound transformer. Counterexample: Prev = 3/2, Prev_l = -4, Prev_u = 2, Prev_L = 3/2, Prev_U = 3/2, Curr = 3/4. Based on the transformer, we get Curr_l = 0, Curr_u = 5/6, Curr_L = 3/4, Curr_U = -5/18, which is a wrong approximation.

```
transformer DeepPoly {
    HardSigmoid ->
        (prev[u] <= -3) ?
            (0, 0, 0, 0) :
        (prev[l] >= 3) ?
            (1, 1, 1, 1) :
        (prev[l] >= -3) ?
            (prev[u] <= 3) ?
                ((prev[l] / 6) + 0.5,
                 (prev[u] / 6) + 0.5,
                 (prev / 6) + 0.5,
                 (prev / 6) + 0.5) :
                ((prev[l] / 6) + 0.5,
                 1,
                 (((0.5 - (prev[l] / 6)) / (prev[u] - prev[l])) * prev)
                   + ((prev[l] / 6) + 0.5)
                   - (((0.5 - (prev[l] / 6)) / (prev[u] - prev[l])) * prev[l]),
                 1) :
        (prev[u] <= 3) ?
            (0,
             (prev[u] / 6) + 0.5,
             0,
```

```
            (((((prev[u] / 6) + 0.5) / (prev[u] - prev[l])) * prev)
              - ((((prev[u] / 6) + 0.5) / (prev[u] - prev[l])) * prev[l]))) :
           (0, 1, 0, 1);
}
```

Listing 3. Sound transformer

Fig. 9. Examples of transformer candidates generated by SAIL. (1) contains syntax errors, (2) is valid but unsound, and (3) is both valid and sound.

## C Validation Semantics

We define the general validation semantics applicable to any programming language. Validation is expressed as a big-step judgment. Given a candidate program $s$, validation either produces a finite set of diagnostics $\mathcal{D}$ or succeeds with a valid abstract syntax tree $t$, where $\mathcal{D}$ aggregates all diagnostics produced during parsing and validation, i.e., $\mathcal{D} = \mathcal{D}_p \cup \mathcal{D}_v$:

$$\frac{\text{Lex}(s) = \vec{\tau} \quad \text{Parse}(\vec{\tau}) = (t, \mathcal{D}_p) \quad \langle \Sigma, \Gamma, \mathcal{M} \rangle \vdash t \Rightarrow \mathcal{D}_v}{\begin{cases} \langle \Sigma, \Gamma, \mathcal{M} \rangle \vdash s \Downarrow \text{err}(\mathcal{D}_p \cup \mathcal{D}_v), & \text{if } \mathcal{D}_p \cup \mathcal{D}_v \neq \emptyset \\ \langle \Sigma, \Gamma, \mathcal{M} \rangle \vdash s \Downarrow \text{ok}(t), & \text{if } \mathcal{D}_p \cup \mathcal{D}_v = \emptyset \end{cases}}$$

Here, $\text{Lex}(s) = \vec{\tau}$ performs lexical analysis on the candidate text $s$, $\text{Parse}(\vec{\tau}) = (t, \mathcal{D}_p)$ constructs its abstract syntax tree (AST) when parsing succeeds, and reports parsing errors by a finite set of diagnostics $\mathcal{D}_p$ otherwise. The judgment $\langle \Sigma, \Gamma, \mathcal{M} \rangle \vdash t \Rightarrow \mathcal{D}_v$ checks the AST for structural and semantic consistency under the symbol table $\Sigma$, typing and shape context $\Gamma$, and metadata map $\mathcal{M}$, producing a set of diagnostics $\mathcal{D}_v$. Validation succeeds iff $\mathcal{D} = \mathcal{D}_p \cup \mathcal{D}_v = \emptyset$. We identify six categories of structural and semantic errors commonly observed in LLM-generated candidates.

(1) Unmatched or missing delimiters. These errors are detected during parsing. The parser ensures that all parentheses, brackets, and braces are properly nested and matched; any violation produces a diagnostic D-DELIM.

(2) Illegal keywords or illegal logical operators. These include errors such as incorrect use of "&&" instead of "`and`" as the conjunction operator. In this instance, if the generated expression fails to pass this test, the validator produces the diagnostic D-OP. These errors can also be tested during parsing.

(3) Improper use of reserved constants or keywords. For instance, the keywords `transformer` and `func` for transformer and function definitions, respectively. The LLMs sometimes incorrectly use `transformer` for function definitions. Reserved keywords must be used in their designated syntactic constructs. If a keyword is used in an invalid context, the validator produces the diagnostic D-RES.

(4) Undefined identifiers or invalid function invocations. Identifiers and calls are checked against the symbol table $\Sigma$:

$$\frac{x \in \text{dom}(\Gamma)}{\langle \Sigma, \Gamma, \mathcal{M} \rangle \vdash x : \Gamma(x)} \text{(T-VAR)} \qquad \frac{f : (\tau_1, \ldots, \tau_n) \rightarrow \tau \in \Sigma \quad \forall i.\ \langle \Sigma, \Gamma, \mathcal{M} \rangle \vdash e_i : \tau_i}{\langle \Sigma, \Gamma, \mathcal{M} \rangle \vdash f(e_1, \ldots, e_n) : \tau} \text{(T-CALL)}$$

where $\Sigma$ is the symbol table mapping identifiers to their declared types, $\Gamma$ is the typing context, $\tau$ denotes the return type of the function, and each $e_i$ is an argument expression that must match the corresponding parameter type $\tau_i$. Violations yield D-ID($x$), where $x$ denotes an undeclared variable or function name. The complete set of rules can be found in [57].

(5) Malformed attribute calls and incorrect metadata indexing. ConstraintFlow uses the square bracket notation (e.g., $e[m]$ where $m$ is a metadata such as $l$ or $L$) to access metadata of a neuron. If the generated candidate incorrectly uses the dot notation (e.g., $e.m$), the validator outputs D-Attr diagnostic.

(6) Type inconsistencies in arithmetic, logical, or element-wise operations. For instance, all logical operators share the same typing pattern. Consider the following type-checking rule where $e \equiv \mathrm{op}(e_1, \ldots, e_n)$:

$$\frac{\mathrm{op} \in \{\mathrm{and}, \mathrm{or}, \mathrm{xor}, \mathrm{not}\} \quad \forall e_i \in \mathrm{args}(\mathrm{op}).\ \langle \Sigma, \Gamma, \mathcal{M} \rangle \vdash e_i : \mathrm{Bool}}{\langle \Sigma, \Gamma, \mathcal{M} \rangle \vdash \mathrm{op}(e_1, \ldots, e_n) : \mathrm{Bool}} \quad \text{(T-LogicOp)}$$

where op denotes a logical operator, $e_i$ are its operands, and $\langle \Sigma, \Gamma, \mathcal{M} \rangle$ represents the symbol table, typing context, and metadata map used for type checking. If any operand fails to have a Boolean type, the validator emits the diagnostic D-Type($e$). The complete set of type-checking rules can be found in [57].

## D Precision Evaluation (Cont.)

We show the complete precision evaluation results for transformers generated by GPT-5 for DeepPoly domain and DeepZ domain.

Table 2. Precision comparison across different networks based on DeepPoly domain.

| Dataset | Network | Training | Activation | Layers | Perturbation $\epsilon$ | Precision | |
|---|---|---|---|---|---|---|---|
| | | | | | | Our work | Handcrafted |
| MNIST | FCN_9×200 | Standard | ReLU | 9 | 0.005 | 0.9691 | 0.9691 |
| | FCN_4×1024 | Standard | ReLU | 4 | 0.005 | 1.0000 | 1.0000 |
| | FCN_6×500 | Standard | ReLU | 6 | 0.005 | 0.9900 | 0.9900 |
| | FCN_6×500 | PGD | ReLU | 6 | 0.005 | 1.0000 | 1.0000 |
| | Conv | DiffAI | ReLU | 3 | 0.005 | 1.0000 | 1.0000 |
| | Conv | PGD | ReLU | 3 | 0.005 | 1.0000 | 1.0000 |
| | Conv | Standard | ReLU | 6 | 0.005 | 1.0000 | 1.0000 |
| | Conv | Standard | ReLU6 | 3 | 0.005 | 1.0000 | ✗ |
| | FCN_3×50 | Standard | ReLU6 | 3 | 0.005 | 0.5100 | ✗ |
| | FCN_3×100 | Standard | ReLU6 | 3 | 0.005 | 0.7300 | ✗ |
| | FCN_4×1024 | Standard | ReLU6 | 4 | 0.005 | 0.9000 | ✗ |
| | FCN_5×100 | DiffAI | ReLU6 | 5 | 0.005 | 0.9000 | ✗ |
| | FCN_6×100 | Standard | ReLU6 | 6 | 0.005 | 0.6465 | ✗ |
| | FCN_6×200 | Standard | ReLU6 | 6 | 0.005 | 0.8384 | ✗ |
| | FCN_6×500 | PGD | ReLU6 | 6 | 0.005 | 1.0000 | ✗ |
| | FCN_6×500 | Standard | ReLU6 | 6 | 0.005 | 0.7879 | ✗ |
| | FCN_9×100 | Standard | ReLU6 | 9 | 0.005 | 0.6800 | ✗ |
| | FCN_9×200 | Standard | ReLU6 | 9 | 0.005 | 0.6768 | ✗ |
| | FCN_3×50 | Standard | HardTanh | 3 | 0.005 | 0.1818 | 0.1818 |
| | FCN_3×100 | Standard | HardTanh | 3 | 0.005 | 0.2323 | 0.2323 |
| | FCN_5×100 | DiffAI | HardTanh | 5 | 0.005 | 0.9500 | 0.9500 |
| | FCN_6×500 | PGD | HardTanh | 6 | 0.005 | 0.5455 | 0.5455 |
| | FCN_3×50 | Standard | HardSigmoid | 3 | 0.005 | 0.2062 | ✗ |
| | FCN_3×100 | Standard | HardSigmoid | 3 | 0.005 | 0.2323 | ✗ |
| | FCN_5×100 | DiffAI | HardSigmoid | 5 | 0.005 | 0.6087 | ✗ |
| | FCN_6×500 | PGD | HardSigmoid | 6 | 0.005 | 1.0000 | ✗ |
| | FCN_9×100 | Standard | HardSigmoid | 9 | 0.005 | 1.0000 | ✗ |

| Dataset | Network | Training | Activation | Layers | Perturbation $\epsilon$ | Precision | |
|---|---|---|---|---|---|---|---|
| | | | | | | Our work | Handcrafted |
| | FCN_9×200 | Standard | HardSigmoid | 9 | 0.005 | 1.0000 | ✗ |
| | FCN_3×50 | Standard | HardSwish | 3 | 0.005 | 0.2653 | ✗ |
| | FCN_3×100 | Standard | HardSwish | 3 | 0.005 | 0.1900 | ✗ |
| | FCN_3×50 | Standard | GELU | 3 | 0.005 | 0.4646 | ✗ |
| | FCN_3×100 | Standard | GELU | 3 | 0.005 | 0.9400 | ✗ |
| | FCN_4×1024 | Standard | GELU | 4 | 0.005 | 1.0000 | ✗ |
| | FCN_5×100 | DiffAI | GELU | 5 | 0.005 | 0.7778 | ✗ |
| | FCN_6×100 | Standard | GELU | 6 | 0.005 | 0.8800 | ✗ |
| | FCN_6×200 | Standard | GELU | 6 | 0.005 | 1.0000 | ✗ |
| | FCN_6×500 | PGD | GELU | 6 | 0.005 | 1.0000 | ✗ |
| | FCN_6×500 | Standard | GELU | 6 | 0.005 | 1.0000 | ✗ |
| | FCN_9×100 | Standard | GELU | 9 | 0.005 | 0.9300 | ✗ |
| | FCN_9×200 | Standard | GELU | 9 | 0.005 | 0.9800 | ✗ |
| | FCN_3×100 | Standard | ELU | 3 | 0.005 | 0.1400 | ✗ |
| CIFAR10 | FCN_4×100 | Standard | ReLU | 4 | 8/255 | 1.0000 | 1.0000 |
| | FCN_6×100 | Standard | ReLU | 6 | 8/255 | 1.0000 | 1.0000 |
| | FCN_9×200 | Standard | ReLU | 9 | 8/255 | 1.0000 | 1.0000 |
| | FCN_7×1024 | Standard | ReLU | 7 | 8/255 | 1.0000 | 1.0000 |
| | Conv | DiffAI | ReLU | 3 | 8/255 | 1.0000 | 1.0000 |
| | Conv | PGD | ReLU | 3 | 8/255 | 1.0000 | 1.0000 |
| | Conv | Standard | ReLU | 3 | 8/255 | 1.0000 | 1.0000 |
| | Conv | Standard | ReLU | 6 | 8/255 | 0.9851 | 0.9851 |
| | Conv | PGD | ReLU | 6 | 8/255 | 1.0000 | 1.0000 |
| | FCN_6×500 | Standard | ReLU | 6 | 8/255 | 1.0000 | 1.0000 |
| | FCN_6×500 | PGD | ReLU | 6 | 8/255 | 1.0000 | 1.0000 |
| | FCN_4×100 | Standard | ReLU6 | 4 | 8/255 | 0.4490 | ✗ |
| | FCN_6×100 | Standard | ReLU6 | 6 | 8/255 | 0.3922 | ✗ |
| | FCN_6×500 | PGD | ReLU6 | 6 | 8/255 | 0.4865 | ✗ |
| | FCN_6×500 | Standard | ReLU6 | 6 | 8/255 | 0.4643 | ✗ |
| | FCN_7×1024 | Standard | ReLU6 | 7 | 8/255 | 0.3077 | ✗ |
| | FCN_9×200 | Standard | ReLU6 | 9 | 8/255 | 0.3721 | ✗ |
| | FCN_4×100 | Standard | HardTanh | 4 | 8/255 | 0.4194 | 0.4194 |
| | FCN_6×100 | Standard | HardTanh | 6 | 8/255 | 0.4474 | 0.4474 |
| | FCN_6×500 | PGD | HardTanh | 6 | 8/255 | 0.3636 | 0.3636 |
| | FCN_6×500 | Standard | HardTanh | 6 | 8/255 | 0.5484 | 0.5484 |
| | FCN_7×1024 | Standard | HardTanh | 7 | 8/255 | 0.9231 | 0.9231 |
| | FCN_9×200 | Standard | HardTanh | 9 | 8/255 | 0.3611 | 0.3611 |
| | FCN_4×100 | Standard | HardSwish | 4 | 8/255 | 0.1154 | ✗ |
| | FCN_4×100 | Standard | HardSigmoid | 4 | 8/255 | 0.3636 | ✗ |
| | FCN_6×500 | PGD | HardSigmoid | 6 | 8/255 | 1.0000 | ✗ |
| | FCN_6×500 | Standard | HardSigmoid | 6 | 8/255 | 0.4894 | ✗ |
| | FCN_7×1024 | Standard | HardSigmoid | 7 | 8/255 | 1.0000 | ✗ |
| | FCN_9×200 | Standard | HardSigmoid | 9 | 8/255 | 1.0000 | ✗ |
| | FCN_4×100 | Standard | GELU | 4 | 8/255 | 0.9630 | ✗ |
| | FCN_6×100 | Standard | GELU | 6 | 8/255 | 0.9649 | ✗ |
| | FCN_6×500 | PGD | GELU | 6 | 8/255 | 1.0000 | ✗ |
| | FCN_6×500 | Standard | GELU | 6 | 8/255 | 0.6000 | ✗ |
| | FCN_7×1024 | Standard | GELU | 7 | 8/255 | 0.9787 | ✗ |

| Dataset | Network | Training | Activation | Layers | Perturbation $\epsilon$ | Precision | |
|---|---|---|---|---|---|---|---|
| | | | | | | Our work | Handcrafted |
| | FCN_9×200 | Standard | GELU | 9 | 8/255 | 0.7358 | ✗ |

Table 3. Precision comparison across different networks based on DeepZ domain.

| Dataset | Network | Training | Activation | Layers | Perturbation $\epsilon$ | Precision | |
|---|---|---|---|---|---|---|---|
| | | | | | | Our work | Handcrafted |
| MNIST | FCN_3×50 | Standard | ReLU | 3 | 0.005 | 0.5204 | 0.5204 |
| | FCN_3×100 | Standard | ReLU | 3 | 0.005 | 0.1122 | 0.1122 |
| | Convolution | Standard | ReLU | 3 | 0.005 | 0.9000 | 0.9000 |
| | Convolution | DiffAI | ReLU | 3 | 0.005 | 1.0000 | 1.0000 |
| | Convolution | PGD | ReLU | 3 | 0.005 | 0.9400 | 0.9400 |
| | Convolution | Standard | ReLU | 6 | 0.005 | 0.8100 | 0.8100 |
| | FCN_3×50 | Standard | HardSigmoid | 3 | 0.005 | 0.2371 | ✗ |
| | FCN_3×100 | Standard | HardSigmoid | 3 | 0.005 | 0.1818 | ✗ |
| | FCN_6×100 | Standard | HardSigmoid | 6 | 0.005 | 0.1531 | ✗ |
| | FCN_6×200 | Standard | HardSigmoid | 6 | 0.005 | 0.2222 | ✗ |
| CIFAR10 | Conv | DiffAI | ReLU | 3 | 8/255 | 1.0000 | 1.0000 |
| | Conv | PGD | ReLU | 3 | 8/255 | 0.8143 | 0.8143 |
| | FCN_6×500 | PGD | ReLU6 | 6 | 8/255 | 0.4595 | ✗ |
| | FCN_6×500 | Standard | ReLU6 | 6 | 8/255 | 0.2500 | ✗ |
| | FCN_4×100 | Standard | HardSigmoid | 4 | 8/255 | 0.3636 | ✗ |
| | FCN_6×500 | PGD | HardSigmoid | 6 | 8/255 | 1.0000 | ✗ |
| | FCN_7×1024 | Standard | HardSigmoid | 7 | 8/255 | 1.0000 | ✗ |
| | FCN_9×200 | Standard | HardSigmoid | 9 | 8/255 | 1.0000 | ✗ |
| | FCN_4×100 | Standard | HardTanh | 4 | 8/255 | 0.3871 | ✗ |
| | FCN_6×500 | PGD | HardTanh | 6 | 8/255 | 0.3636 | ✗ |
| | FCN_6×500 | Standard | HardTanh | 6 | 8/255 | 0.3871 | ✗ |
| | FCN_7×1024 | Standard | HardTanh | 7 | 8/255 | 0.8462 | ✗ |

## E Computation Cost Evaluation

Table 4 reports both the runtime of the best-performing synthesis using GPT-5 for synthesizing common abstract transformers in the DeepPoly domain, and the corresponding average verification cost (in seconds) on a standard CIFAR-10 benchmark, evaluated on two fully connected networks: a 4-layer network with 100 neurons per layer and a 9-layer network with 200 neurons per layer.

Table 4. Synthesis and verification cost (in seconds) of common operators.

| Operator | ReLU | ReLU6 | HardSigmoid | HardSwish | HardTanh |
|---|---|---|---|---|---|
| Synthesis Cost | 8.56 | 407.79 | 297.40 | 451.92 | 209.80 |
| Verification Cost (FCN 4×100) | 6.6304 | 12.6875 | 6.2336 | 5.8500 | 14.7250 |
| Verification Cost (FCN 9×200) | 32.3159 | 48.8187 | 14.9709 | 42.1219 | 50.9219 |

## F Sound Abstract Transformer for ELU Activation

Fig. 10 visualizes the synthesized sound abstract transformer for ELU activation in the DeepPoly Domain.

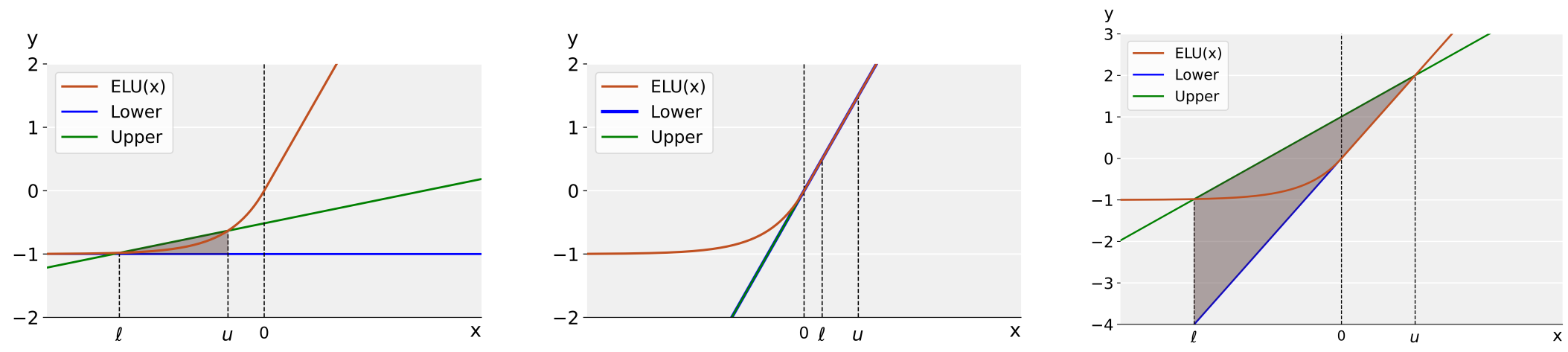

(a) $l < u < 0$. (b) $0 < l < u$. (c) $l < 0 < u$.

Fig. 10. **DeepPoly Transformer for ELU.** The transformer is divided into three cases: (1) for $l < u < 0$, the lower and upper bounds are $y = -1$ and the secant line connecting $(l, \mathrm{ELU}(l))$ and $(u, \mathrm{ELU}(u))$, respectively; (2) for $0 < l < u$, the lower and upper bounds are $y = x$ and the same secant line; (3) for the mixed case $(l < 0 < u)$, the upper bound is again the secant through $(l, \mathrm{ELU}(l))$ and $(u, \mathrm{ELU}(u))$, while the lower bound is $y = x$. Since $\mathrm{ELU}(x)$ is monotonic, the scalar bounds correspond directly to $\mathrm{ELU}(l)$ and $\mathrm{ELU}(u)$.

## G Synthesizing Abstract Transformer For Differentiable Programs

Fig. 11 visualizes the synthesized transformer for the first-order derivative of the Sigmoid function, i.e., $\sigma'(x) = \sigma(x)(1 - \sigma(x))$, in the interval domain.

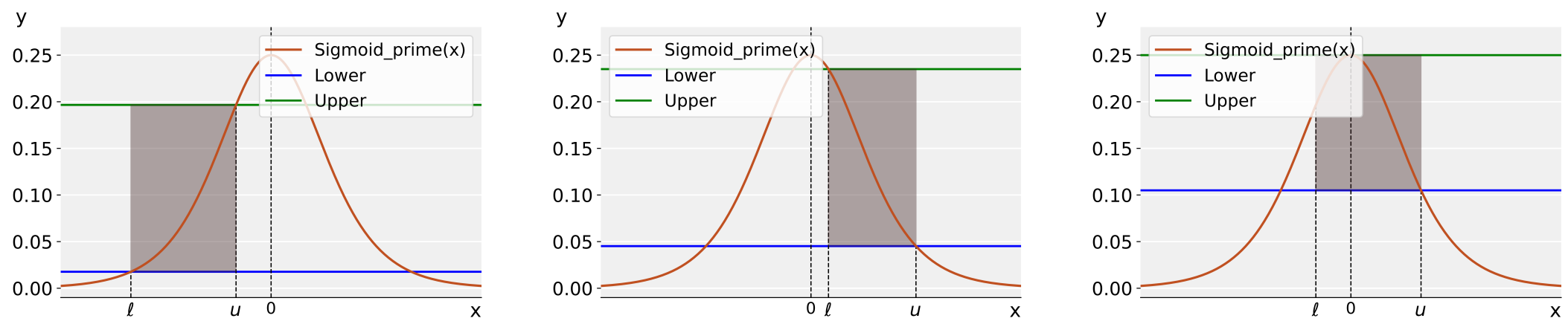

(a) $l < u < 0$. (b) $0 < l < u$. (c) $l < 0 < u$.

Fig. 11. **Transformer for the first-order derivative of the Sigmoid function in the interval domain.** The transformer is divided into three cases: (1) for $l < u < 0$, the lower and upper scalar bounds are $\sigma(l)(1 - \sigma(l))$ and $\sigma(u)(1 - \sigma(u))$ respectively; (2) for $0 < l < u$, the scalar bounds are $\sigma(u)(1 - \sigma(u))$ and $\sigma(l)(1 - \sigma(l))$ respectively; (3) for the mixed case $(l < 0 < u)$, the upper scalar bound is 0.25, and the lower bound is $min(\sigma(l)(1 - \sigma(l)), \sigma(u)(1 - \sigma(u)))$. In the subfigure, we use the case when $u > -l$ as an example.

## H Fallback Abstract Transformers

We provide trivially sound abstract transformers as a fallback when the synthesis fails within the maximum number of rounds. Specifically, we use the $\top$ element of the abstract domain as a general fallback and employ endpoint-based bounds for certain monotonic functions (e.g., HardSigmoid, HardTanh, HardSwish, ReLU, ReLU6). We list all the fallbacks for supported operators used in the experiment in the Interval, DeepPoly, and DeepZ domains, expressed in the DSL of ConstraintFlow, as shown in Fig. 12.

```
transformer ibp {
    Abs -> (-inf, inf);
}

transformer ibp {
    Neuron_add -> (-inf, inf);
}

transformer ibp {
    Affine -> (-inf, inf);
}

transformer ibp {
    Avgpool -> (-inf, inf);
}

func hsm(Float x) = x<-3 ? 0 : (x<3 ? x/6+0.5 : 1);
transformer ibp {
    HardSigmoid -> (hsm(prev[l]), hsm(prev[u]));
}

func hsw(Float x) = x<-3 ? 0 : (x<3 ? x*(x+3)/6 : x);
transformer ibp {
    HardSwish -> (hsw(prev[l]), hsw(prev[u]));
}

func ht(Float x) = x<-1 ? -1 : (x<1 ? x : 1);
transformer ibp {
    HardTanh -> (ht(prev[l]), ht(prev[u]));
}

transformer ibp {
    Neuron_max -> (-inf, inf);
}

transformer ibp {
    Maxpool -> (-inf, inf);
}

transformer ibp {
    Neuron_min -> (-inf, inf);
}

transformer ibp {
    Minpool -> (-inf, inf);
}

transformer ibp {
    Neuron_mult -> (-inf, inf);
}

func hl(Float x) = x<0 ? 0 : x;
```

```
transformer ibp {
    Relu -> (hl(prev[l]), hl(prev[u]));
}

func hl6(Float x) = x<0 ? 0 : (x<6 ? x : 6);
transformer ibp {
    Relu6 -> (hl6(prev[l]), hl6(prev[u]));
}

transformer ibp {
    Gelu -> (-inf, inf);
}

transformer ibp {
    Elu -> (-inf, inf);
}

transformer ibp {
    Sigmoid -> (-inf, inf);
}
```

Listing 1. Fallback transformers for all supported operators used in the experiments in the Interval domain, expressed in DSL.

```
transformer deeppoly {
    Abs -> (-inf, inf, -inf, inf);
}

transformer deeppoly {
    Neuron_add -> (-inf, inf, -inf, inf);
}

transformer deeppoly {
    Affine -> (-inf, inf, -inf, inf);
}

transformer deeppoly {
    Avgpool -> (-inf, inf, -inf, inf);
}

func hsm(Float x) = x<-3 ? 0 : (x<3 ? x/6+0.5 : 1);
transformer deeppoly {
    HardSigmoid -> (hsm(prev[l]), hsm(prev[u]), hsm(prev[l]), hsm(prev[u]));
}

func hsw(Float x) = x<-3 ? 0 : (x<3 ? x*(x+3)/6 : x);
transformer deeppoly {
    HardSwish -> (hsw(prev[l]), hsw(prev[u]), hsw(prev[l]), hsw(prev[u]));
}

func ht(Float x) = x<-1 ? -1 : (x<1 ? x : 1);
transformer deeppoly {
    HardTanh -> (ht(prev[l]), ht(prev[u]), ht(prev[l]), ht(prev[u]));
```

```
}

transformer deeppoly {
    Neuron_max -> (-inf, inf, -inf, inf);
}

transformer deeppoly {
    Maxpool -> (-inf, inf, -inf, inf);
}

transformer deeppoly {
    Neuron_min -> (-inf, inf, -inf, inf);
}

transformer deeppoly {
    Minpool -> (-inf, inf, -inf, inf);
}

transformer deeppoly {
    Neuron_mult -> (-inf, inf, -inf, inf);
}

func hl(Float x) = x<0 ? 0 : x;
transformer deeppoly {
    Relu -> (hl(prev[l]), hl(prev[u]), hl(prev[l]), hl(prev[u]));
}

func hl6(Float x) = x<0 ? 0 : (x<6 ? x : 6);
transformer deeppoly {
    Relu6 -> (hl6(prev[l]), hl6(prev[u]), hl6(prev[l]), hl6(prev[u]));
}

transformer deeppoly {
    Gelu -> (-inf, inf, -inf, inf);
}

transformer deeppoly {
    Elu -> (-inf, inf, -inf, inf);
}

transformer deeppoly {
    Sigmoid -> (-inf, inf, -inf, inf);
}
```

Listing 2. Fallback transformers for all supported operators used in the experiments in the DeepPoly domain, expressed in DSL.

```
transformer deepz {
    Abs -> (-inf, inf, inf * eps);
}

transformer deepz {
    Neuron_add -> (-inf, inf, inf * eps);
```

```
}

transformer deepz {
    Affine -> (-inf, inf, inf * eps);
}

transformer deepz {
    Avgpool -> (-inf, inf, inf * eps);
}

func hsm(Float x) = x<-3 ? 0 : (x<3 ? x/6+0.5 : 1);
transformer deepz {
    HardSigmoid -> (hsm(prev[l]), hsm(prev[u]), (hsm(prev[l])+hsm(prev[u]))/2 + (hsm(prev[u])-hsm(prev[l]))/2 * eps);
}

func hsw(Float x) = x<-3 ? 0 : (x<3 ? x*(x+3)/6 : x);
transformer deepz {
    HardSwish -> (hsw(prev[l]), hsw(prev[u]), (hsw(prev[l])+hsw(prev[u]))/2 + (hsw(prev[u])-hsw(prev[l]))/2 * eps);
}

func ht(Float x) = x<-1 ? -1 : (x<1 ? x : 1);
transformer deepz {
    HardTanh -> (ht(prev[l]), ht(prev[u]), (ht(prev[l])+ht(prev[u]))/2 + (ht(prev[u])-ht(prev[l]))/2 * eps);
}

transformer deepz {
    Neuron_max -> (-inf, inf, inf * eps);
}

transformer deepz {
    Maxpool -> (-inf, inf, inf * eps);
}

transformer deepz {
    Neuron_min -> (-inf, inf * eps);
}

transformer deepz {
    Minpool -> (-inf, inf, inf * eps);
}

transformer deepz {
    Neuron_mult -> (-inf, inf, inf * eps);
}

func hl(Float x) = x<0 ? 0 : x;
transformer deepz {
    Relu -> (hl(prev[l]), hl(prev[u]), (hl(prev[l])+hl(prev[u]))/2 + (hl(prev[u])-hl(prev[l]))/2 * eps);
```

```
}

func hl6(Float x) = x<0 ? 0 : (x<6 ? x : 6);
transformer deepz {
    Relu6 -> (hl6(prev[l]), hl6(prev[u]), (hl6(prev[l])+hl6(prev[u]))/2 + (hl6(prev[u])-hl6(
     prev[l]))/2 * eps);
}

transformer deepz {
    Gelu -> (-inf, inf, inf * eps);
}

transformer deepz {
    Elu -> (-inf, inf, inf * eps);
}

transformer deepz {
    Sigmoid -> (-inf, inf, inf * eps);
}
```

Listing 3. Fallback transformers for all supported operators used in the experiments in the DeepZ domain, expressed in DSL.

Fig. 12. Fallback abstract transformers used when synthesis does not converge. (1) shows the fallbacks in the Interval domain, (2) shows the fallbacks in the DeepPoly domain, and (3) shows the fallbacks in the DeepZ domain, all expressed in DSL.

## I Ablation Study (Cont.)

Fig. 13, Fig. 14 and Fig. 15 show the effect of validation-repair module and cost-function guidance on DeepPoly transformer synthesis using Llama4-Maverick.

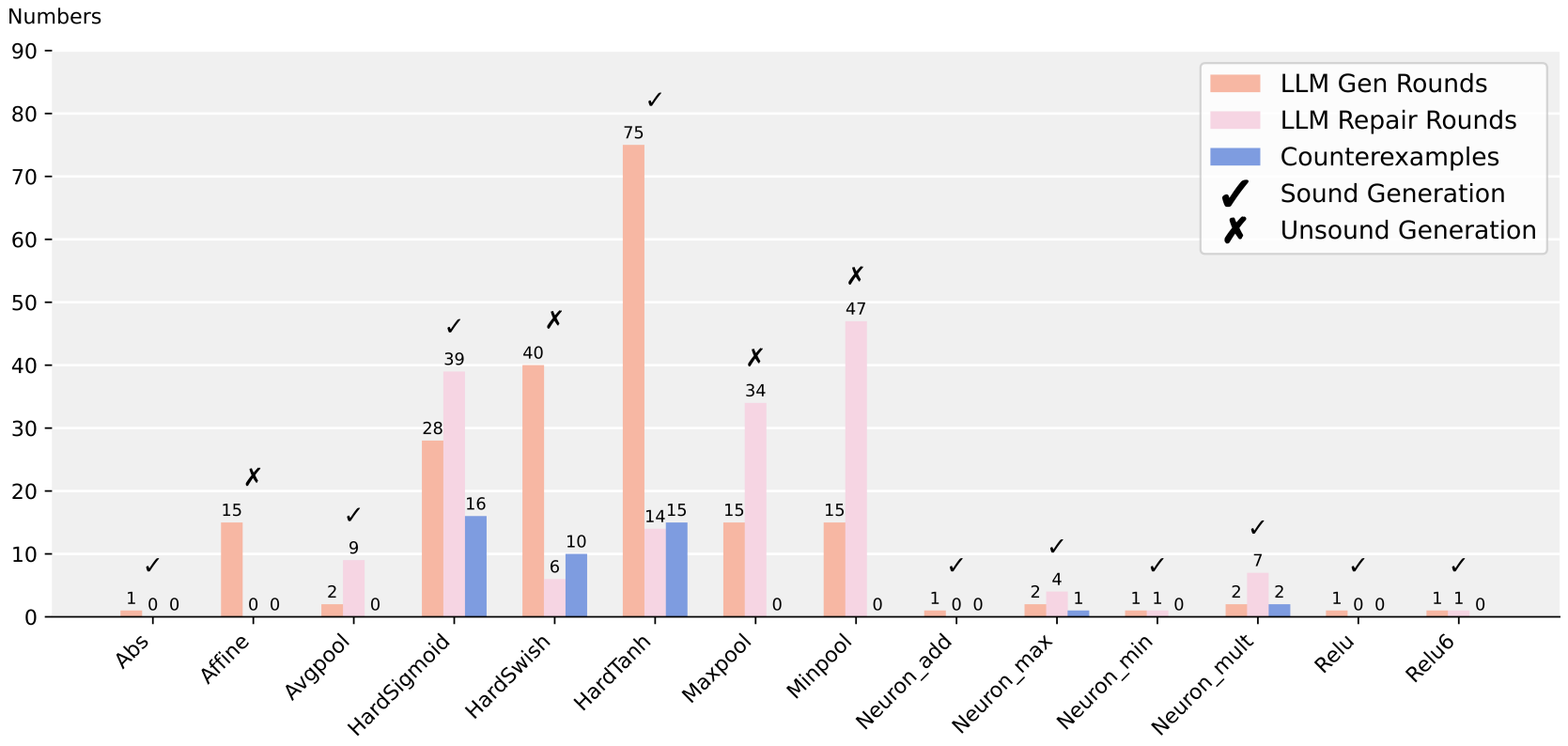

Fig. 13. With cost-function guidance. The model converges to sound transformers within a few refinement rounds.

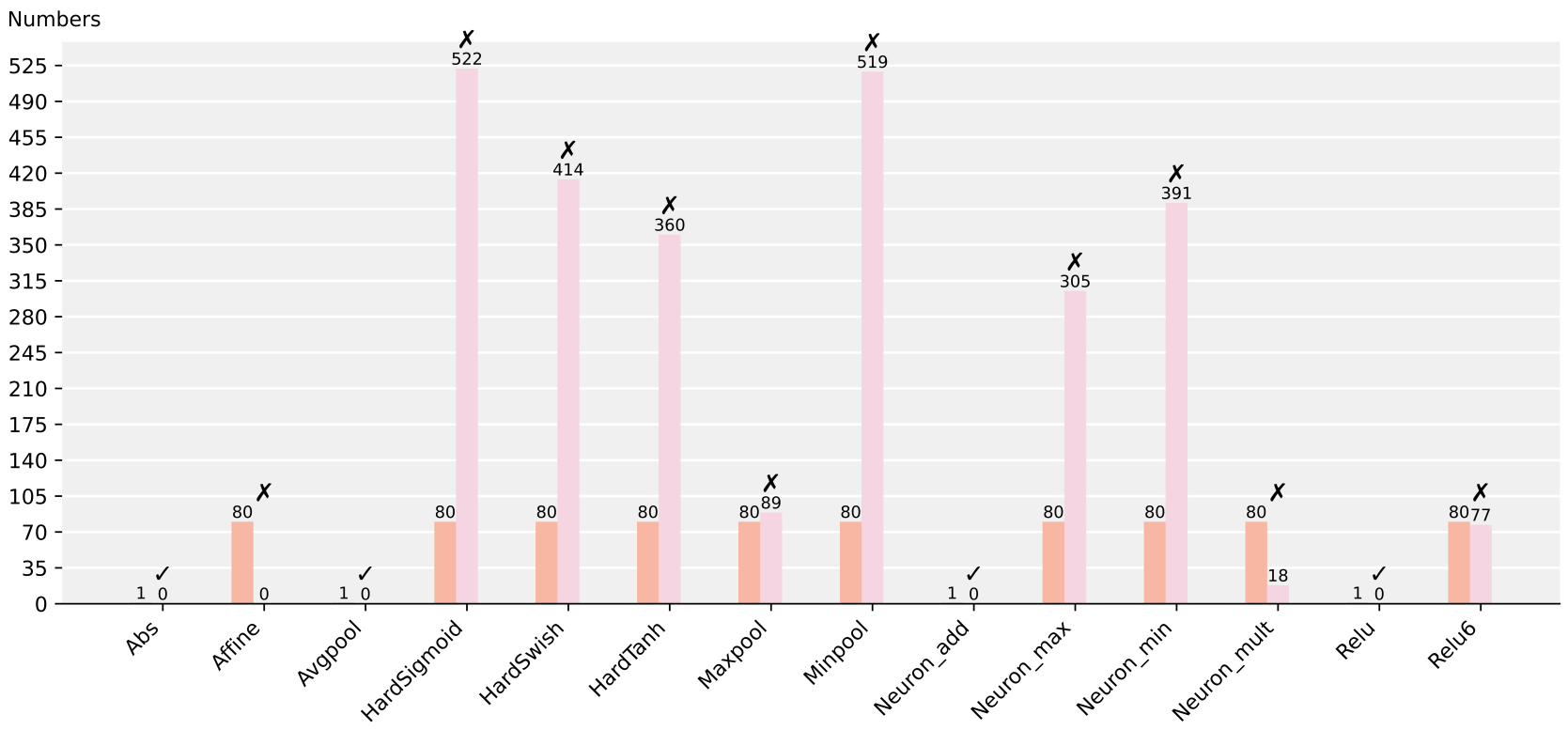

Fig. 14. Without cost-function guidance but with repair module. The model is able to produce syntactically and semantically valid transformers, but they are unsound in most cases.

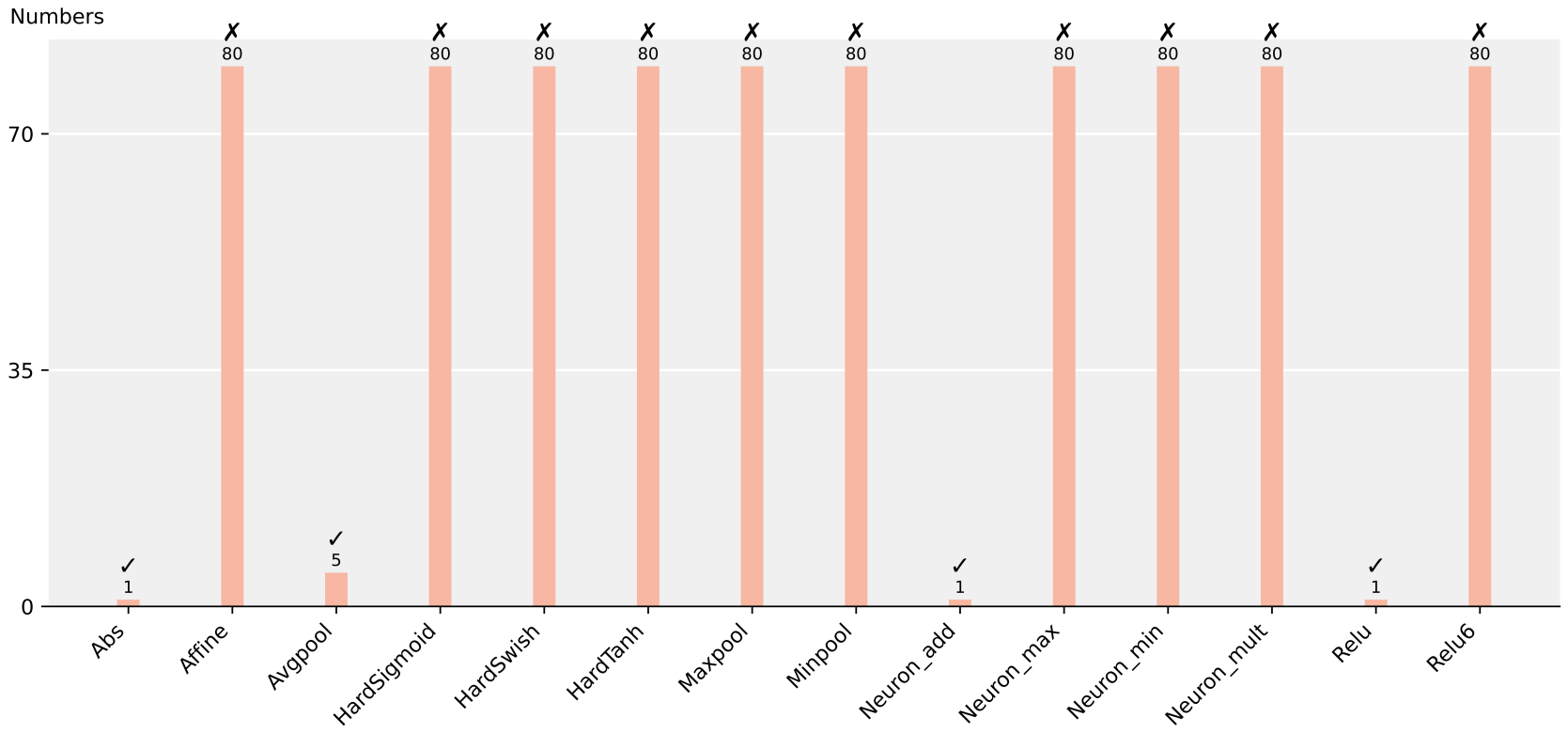

Fig. 15. Without repair module and cost-function guidance. The model often produces syntactically invalid and unsound transformers.

## J Performance of Sound Abstract Interpreters Synthesis across Multiples Models and Domains

Fig. 16, Fig. 17, Fig. 18, Fig. 19, Fig. 20, and Fig. 21 show the process of GPT-5, Llama4-Maverick, Claude-Opus-4 synthesizing multiple transformers across DeepPoly, DeepZ and Interval domain.

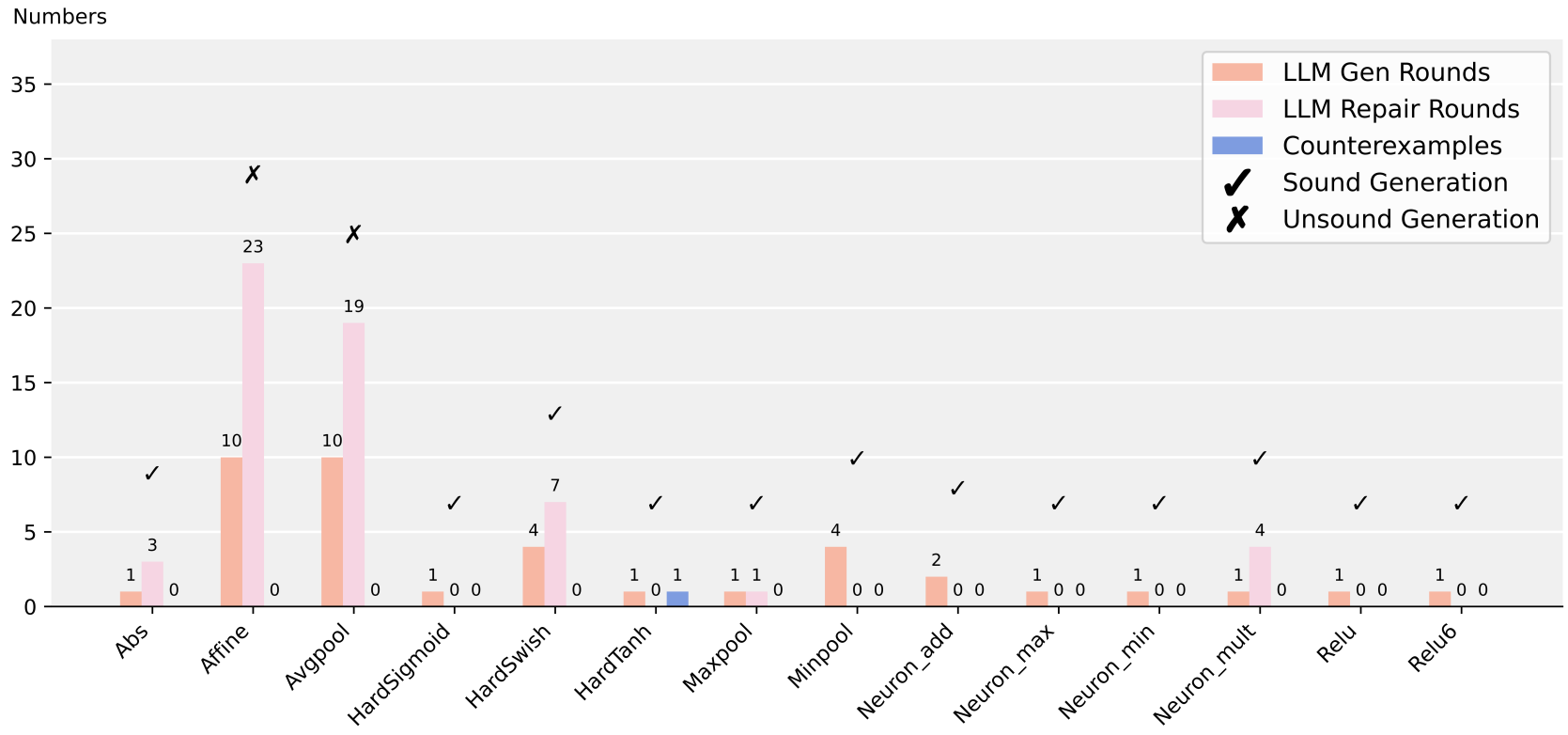

Fig. 16. Quantitative results of GPT-5 generating transformers under the DeepPoly domain.

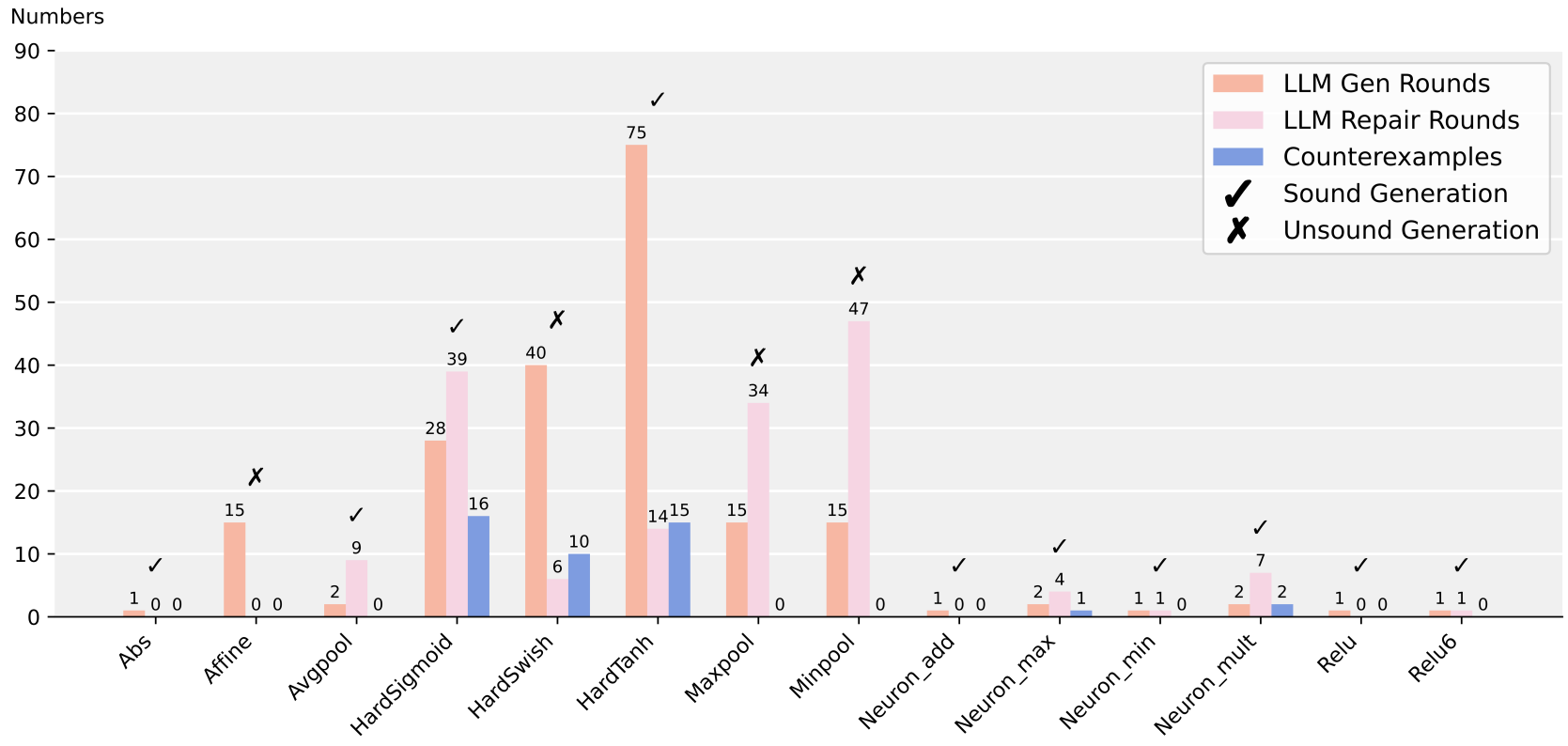

Fig. 17. Quantitative results of Llama4-Maverick generating transformers under the DeepPoly domain.

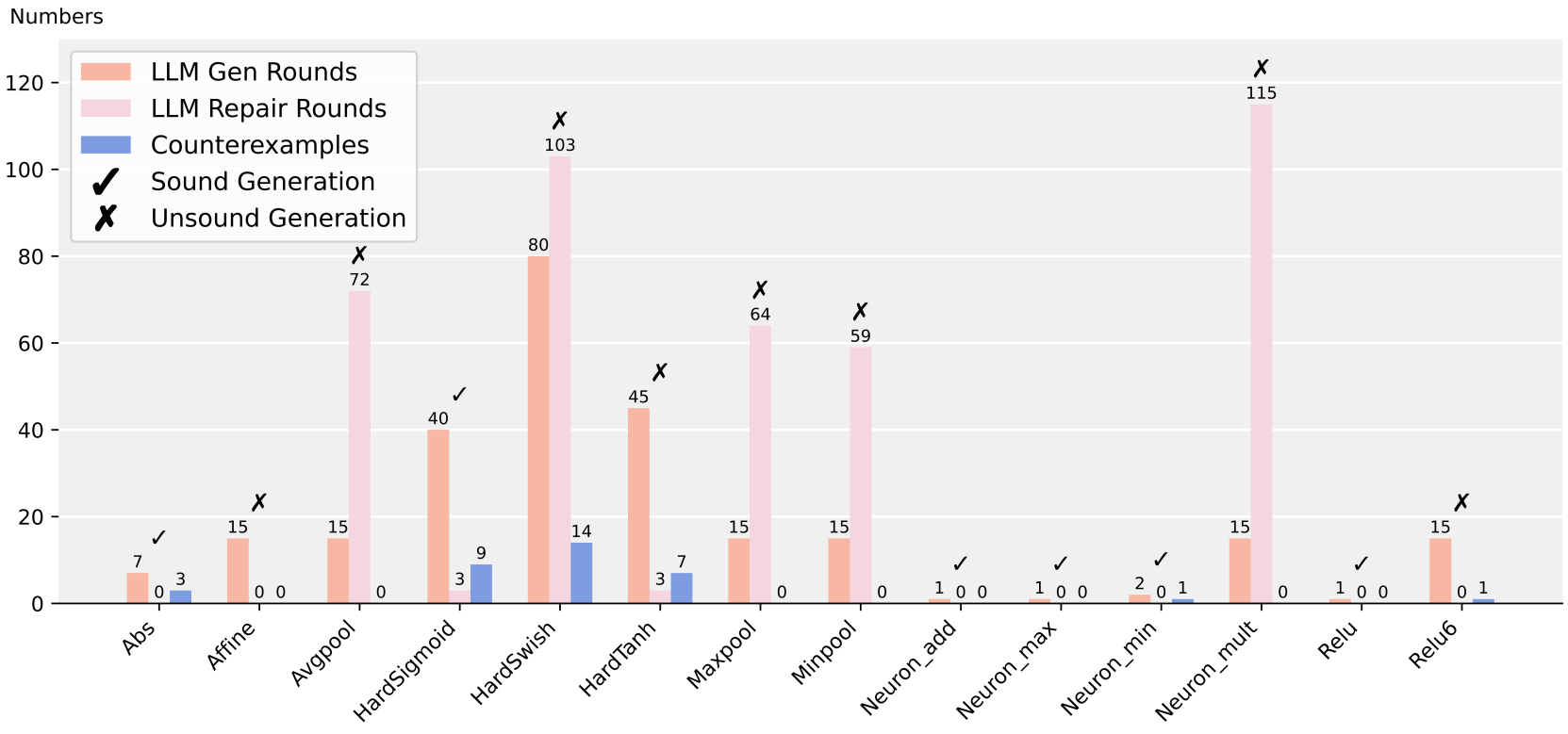

Fig. 18. Quantitative results of Claude-Opus–4 generating transformers under the DeepPoly domain.

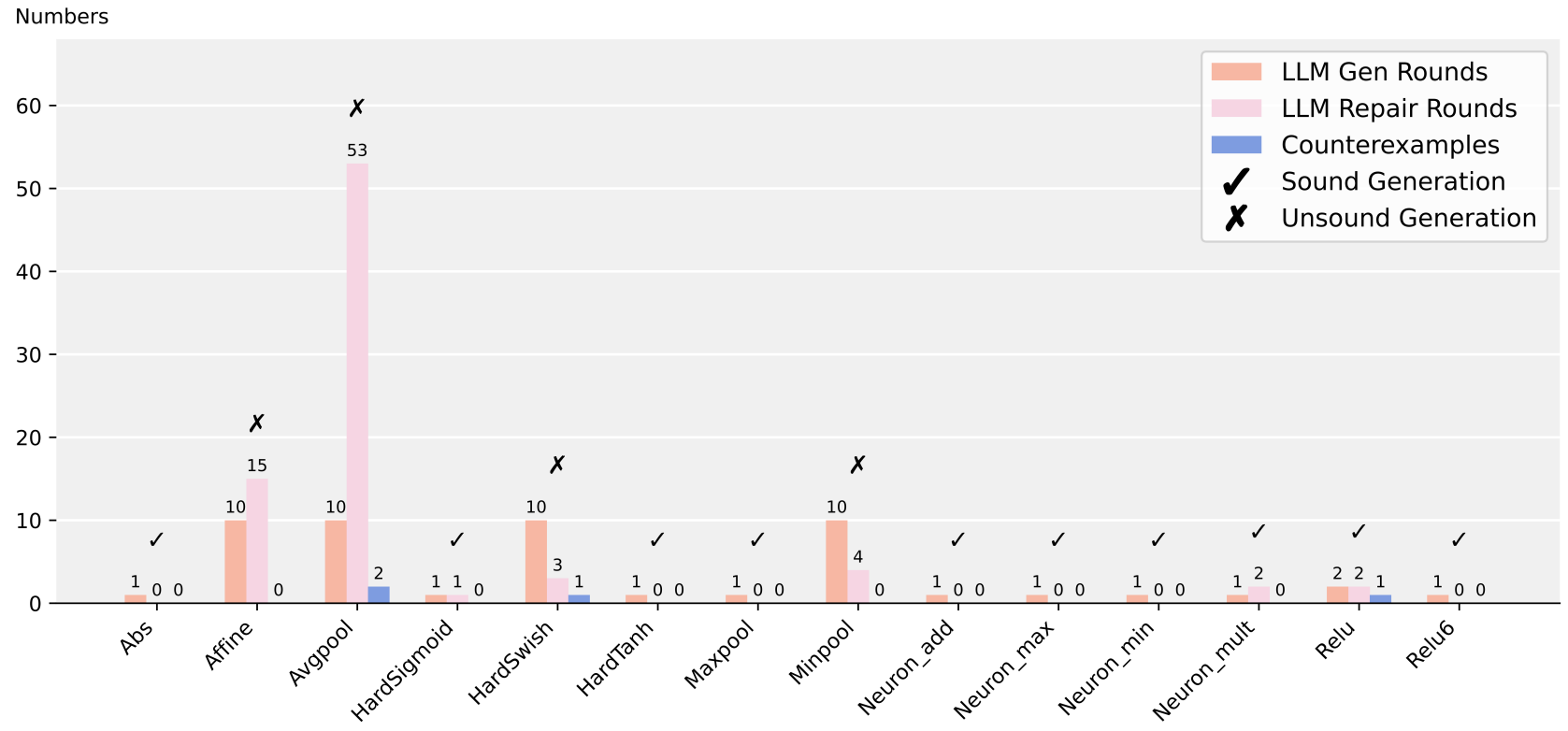

Fig. 19. Quantitative results of GPT-5 generating transformers under the DeepZ domain.

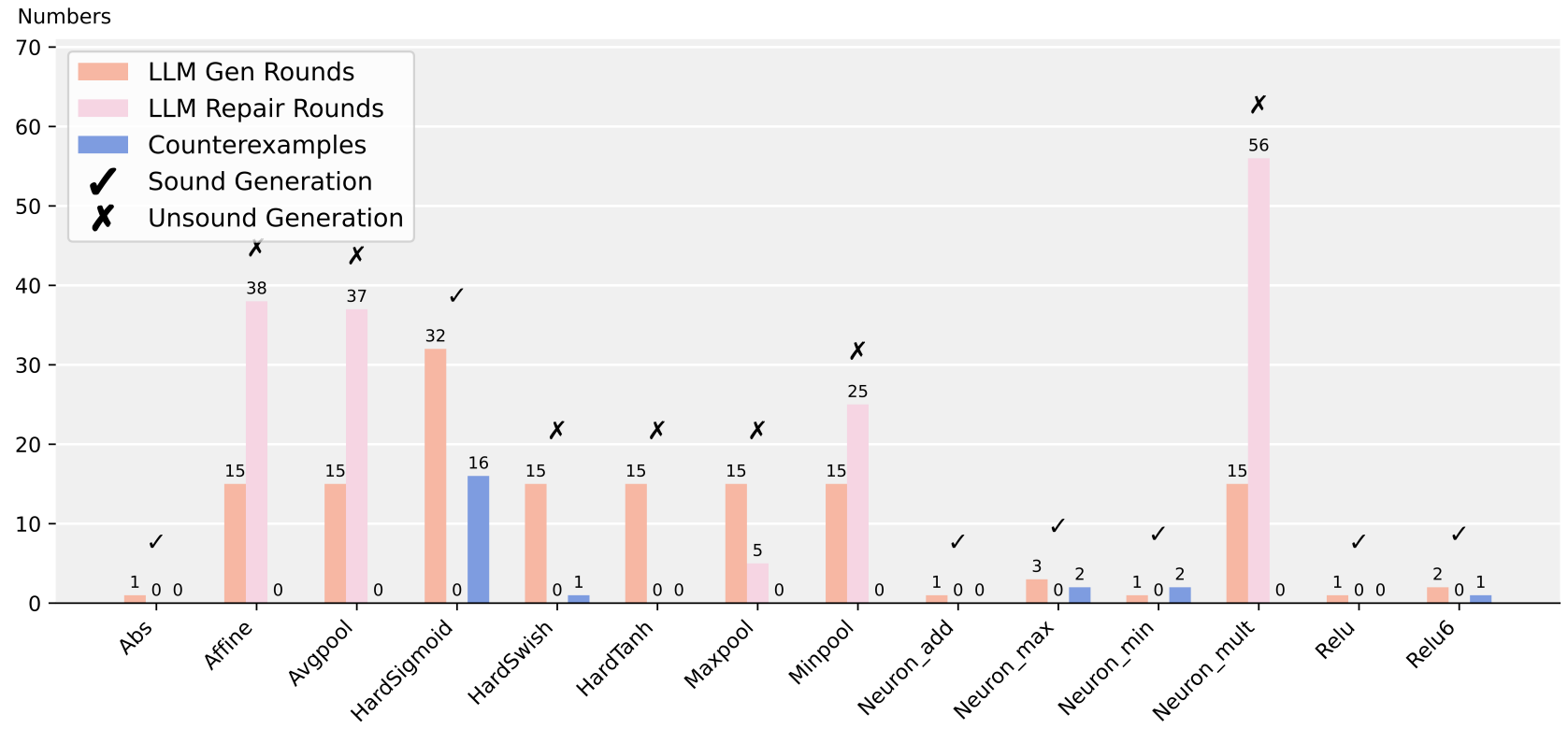

Fig. 20. Quantitative results of Claude-Opus-4 generating transformers under the DeepZ domain.

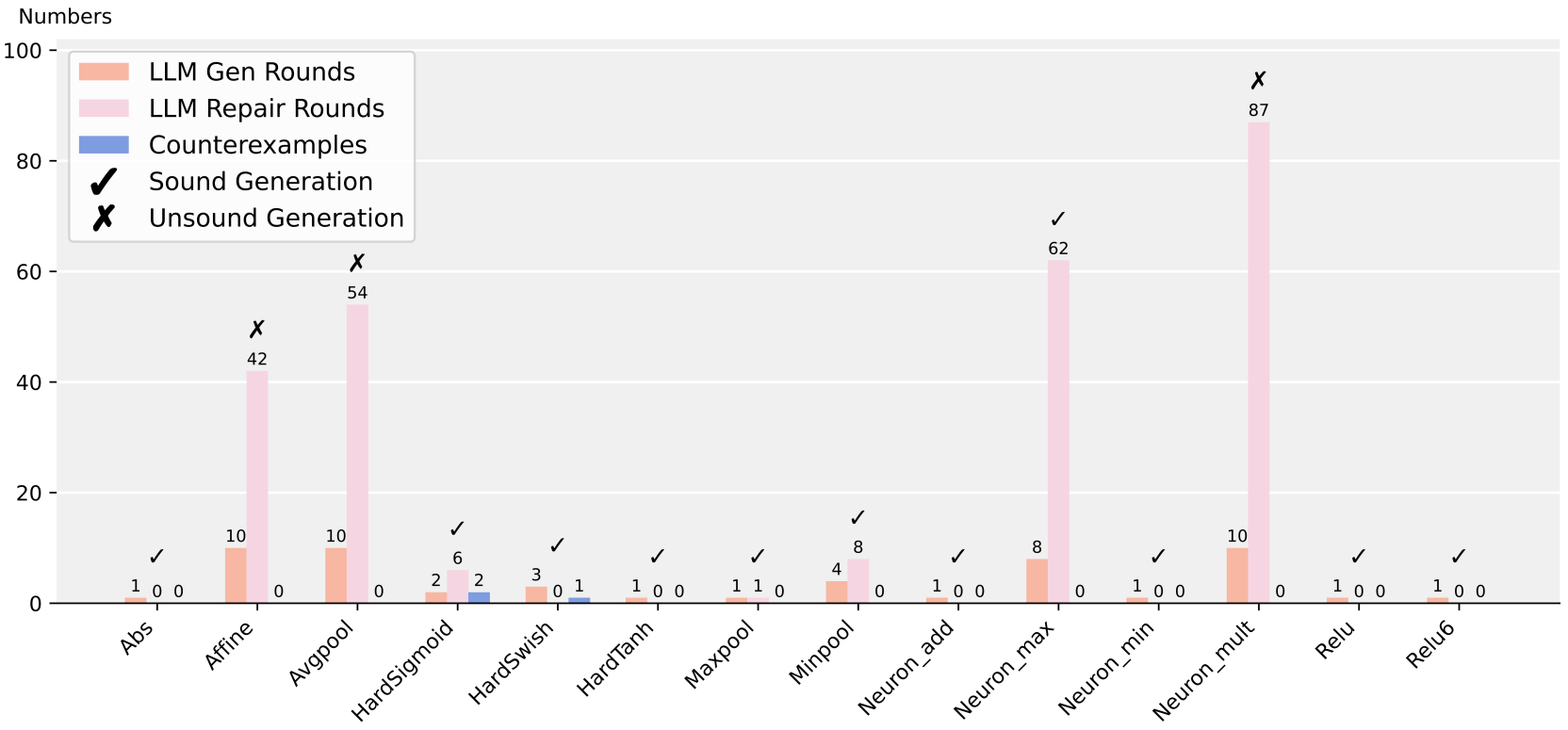

Fig. 21. Quantitative results of GPT5 generating transformers under the Interval domain.